\documentclass[12pt]{article}

\usepackage{amsmath,amssymb,amsfonts,amsthm,hyperref}
\usepackage{graphicx,multirow,caption,bbm,latexsym,epsfig}

\newcommand{\bs}{\begin{subequations}}
\newcommand{\es}{\end{subequations}}
\newcommand{\be}{\begin{equation}}
\newcommand{\ee}{\end{equation}}
\newcommand{\ba}{\begin{eqnarray}}
\newcommand{\ea}{\end{eqnarray}}

\allowdisplaybreaks

\begin{document}

\title{
\normalsize \hfill CFTP/19-013
\\[6mm]
\LARGE No strong $CP$ violation up to the one-loop level
in a two-Higgs-doublet model}

\author{
  \addtocounter{footnote}{2}
  P.~M.~Ferreira$^{(1,2)}$\thanks{\tt pmmferreira@fc.ul.pt}
  \ {\normalsize and}
  L.~Lavoura$^{(3)}$\thanks{\tt balio@cftp.tecnico.ulisboa.pt}
  \\*[3mm]
  $^{(1)}\!$
  \small Instituto~Superior~de~Engenharia~de~Lisboa~---~ISEL,
  1959-007~Lisboa, Portugal
  \\[2mm]
  $^{(2)}\!$
  \small Centro~de~F\'\i sica~Te\'orica~e~Computacional,
  Faculdade~de~Ci\^encias, Universidade~de~Lisboa, \\
  \small Av.~Prof.~Gama~Pinto~2, 1649-003~Lisboa, Portugal
  \\[2mm]
  $^{(3)}\!$
  \small Universidade de Lisboa, Instituto Superior T\'ecnico, CFTP, \\
  \small Av.~Rovisco~Pais~1, 1049-001~Lisboa, Portugal
  \\*[2mm]
}

\date{\today}

\maketitle

\begin{abstract}
  We put forward a two-Higgs-doublet model,
  furnished with a $\mathbbm{Z}_3$ symmetry,
  wherein $CP$ is conserved in the dimension-four terms of the Lagrangian
  and is softly broken in the scalar potential.
  The new particles of our model are one neutral scalar $H$,
  one neutral pseudoscalar $A$,
  and two charged scalars $H^\pm$.
  In our model the only locus of $CP$ violation is the CKM matrix.
  Strong $CP$ violation is absent both at the tree and one-loop levels.
  We work out the phenomenological constraints on our model,
  which features flavour-changing neutral Yukawa interactions,
  showing that the new scalar particles may in some cases
  be lighter than 500\,GeV.
\end{abstract}

\newpage

\section{Introduction}

Non-perturbative effects in Quantum Chromodynamics (QCD)
may lead to $P$ and $CP$ violation,
characterized by a parameter $\theta$,
in hadronic processes.
The experimental upper bound on the electric dipole moment of the neutron
necessitates $\theta \lesssim 10^{-9}$.\footnote{For a recente estimate
  of the maximum possible value of $\theta$,
  see ref.~\cite{Dragos:2019oxn}.}
The presence in the Lagrangian of this unnaturally small parameter
is known as the `strong $CP$ problem'.

The angle $\theta$ is the sum of two terms,
$\theta_\mathrm{QCD}$ and $\theta_\mathrm{QFD}$.
Here,
$\theta_\mathrm{QCD}$ is the value of a $P$-
and $CP$-violating angle in the QCD vacuum,
and $\theta_\mathrm{QFD}$ originates in the chiral rotation
of the quark fields needed to render the quark masses real and positive.
Let $p$ and $n$ denote the three up-type quarks
and the three down-type quarks,
respectively,
in a weak basis.
Let the mass terms of those quarks be given by
\be
\mathcal{L}_\mathrm{mass} =
- \bar p_L M_p p_R - \bar n_L M_n n_R + \mathrm{H.c.},
\ee
where $M_p$ and $M_n$ are $3 \times 3$ matrices in flavour space.
Then,
$\theta_\mathrm{QFD} = \arg{\det{\left( M_p M_n \right)}}$.

There are two general approaches to solving the strong $CP$ problem.
In the first approach it is claimed that $\theta$
has no significance or physical consequences;
theories with different values of $\theta$ are equivalent
and one may set $\theta$ to zero without loss of generality.
This may happen either because one of the quarks
is massless\footnote{For a recent speculation that some quarks
  may be massless,
  see ref.~\cite{Iwazaki:2019hdz}.}
or because of the presence in the theory
of a Peccei--Quinn symmetry~\cite{PQ};
there are also claims that QCD dynamics itself
cures the strong $CP$ problem.\footnote{For a recent instance of such a claim,
  see ref.~\cite{Bardeen:2018fej}.}
The second approach,
which we shall follow,
acknowledges the strong $CP$ problem and
tries to find some symmetry that naturally leads
to the smallness of $\theta$.
One firstly assumes the dimension-four part of the Lagrangian
to be either $CP$-symmetric or $P$-symmetric;
this assumption sets $\theta_\mathrm{QCD}$ to zero.
The $CP$ or $P$ symmetry must be either softly or spontaneously broken;
one performs this breaking in such a way that $\theta_\mathrm{QFD}$
turns out to be zero at the tree level,
because of some peculiar form of $M_p$ and $M_n$.
Still,
it is difficult to avoid loop contributions to $\theta_\mathrm{QFD}$
arising from the quark self-energies $\Sigma$;
they add to the tree-level mass matrices $M$ and then
\be
\arg{\det{\left( M + \Sigma \right)}}
\approx \mathrm{Im} \left[ \mathrm{tr} \left( M^{-1} \Sigma \right) \right]
\ee
is in general nonzero.
Artful models are able to obtain
$\mathrm{Im} \left[ \mathrm{tr} \left( M_p^{-1} \Sigma_p \right) \right]
+ \mathrm{Im} \left[ \mathrm{tr} \left( M_n^{-1} \Sigma_n \right) \right]$
equal to zero at the one-loop level and sometimes even at the two-loop level.

There are various ways to achieve quark mass matrices
displaying $\arg{\det{\left( M_p M_n \right)}} = 0$.
Most of those ways,
collectively known as Barr--Nelson-type models,
employ extra quarks.
There are also many models for solving the strong $CP$ problem
that use extra gauge symmetries,
especially the left--right symmetry
$SU(2)_\mathrm{L} \times SU(2)_\mathrm{R} \times U(1)_{B-L}$.
In this paper we propose a simple extension of the Standard Model (SM),
with gauge group $SU(2)_\mathrm{L} \times U(1)$ and without any extra fermions,
that partially solves the strong $CP$ problem.
Our model is a two-Higgs-doublet model (2HDM)~\cite{2HDM}.

In a 2HDM the quark Yukawa Lagrangian is
\be
\mathcal{L}_\mathrm{Yukawa} = - \sum_{j,k=1}^3 \sum_{a=1}^2
\bar Q_{Lj} \left[ \Phi_a \left( \Gamma_a \right)_{jk}
  n_{Rk} + \tilde \Phi_a \left( \Delta_a \right)_{jk} p_{Rk}
  \right] + \mathrm{H.c.}
\ee
where
$\Phi_a = \left( \begin{array}{cc} \phi_a^+, & \phi_a^0 \end{array} \right)^T$
and $\tilde \Phi_a = \left( \begin{array}{cc}
  {\phi_a^0}^\ast, & - \phi_a^- \end{array} \right)^T$ for $a = 1, 2$
are scalar doublets of $SU(2)_\mathrm{L}$.
Furthermore,
$\bar Q_{Lj} = \left( \begin{array}{cc}
  \bar p_{Lj}, & \bar n_{Lj} \end{array} \right)$,
and the $\Gamma_a$ and $\Delta_a$ are four $3 \times 3$ matrices
in flavour space containing the Yukawa coupling constants.
We expand the scalar doublets as
\be
\label{fmldpdi}
\Phi_a = e^{i \aleph_a} \left( \begin{array}{c} \phi_a^+ \\
  \left( v_a + \rho_a + i \eta_a \right) \left/ \sqrt{2} \right.
  \end{array} \right),
\ee
where $v_a \exp{\left( i \aleph_a \right)} \left/ \sqrt{2} \right.
= \left\langle 0 \left| \phi_a^0 \right| 0 \right\rangle$
and the $v_a$ are non-negative real by definition.
We define $v = \sqrt{v_1^2 + v_2^2} = 2 m_W / g = 246$\,GeV
and $\tan{\beta} \equiv v_2 / v_1$;
then,
\be
\frac{v_1}{v} = \cos{\beta}, \quad \frac{v_2}{v} = \sin{\beta},
\ee
where the angle $\beta$ is in the first quadrant.
There is one physical pseudoscalar $A$
and one unphysical (Goldstone boson) pseudoscalar $G^0$:
\be
\left( \begin{array}{c} G^0 \\ A \end{array} \right) =
\left( \begin{array}{cc} c_\beta & s_\beta \\
  s_\beta & - c_\beta \end{array} \right)
\left( \begin{array}{c} \eta_1 \\ \eta_2 \end{array} \right).
\ee
(From now on,
$s_\xi \equiv \sin{\xi}$ and $c_\xi \equiv \cos{\xi}$
for any needed angle $\xi$.)
There is a pair of physical charged scalars $H^\pm$
and a pair of unphysical (Goldstone bosons) charged scalars $G^\pm$:
\be
\left( \begin{array}{c} G^\pm \\ H^\pm \end{array} \right) =
\left( \begin{array}{cc} c_\beta & s_\beta \\
  s_\beta & - c_\beta \end{array} \right)
\left( \begin{array}{c} \phi_1^\pm \\ \phi_2^\pm \end{array} \right).
\ee
There are two physical neutral scalars $h$ and $H$:
\be
\left( \begin{array}{c} h \\ H \end{array} \right) =
\left( \begin{array}{cc} s_\alpha & - c_\alpha \\
  - c_\alpha & - s_\alpha \end{array} \right)
\left( \begin{array}{c} \rho_1 \\ \rho_2 \end{array} \right).
\ee
The neutral scalar $h$ is chosen to coincide with the LHC-observed particle
with mass 125\,GeV.
In our specific 2HDM
there are softly-broken $CP$ and $\mathbbm{Z}_3$ symmetries
such that $h$ and $H$ do \emph{not}\/ mix with $A$.
The interaction Lagrangian between a scalar and
a pair of gauge bosons is
\bs
\ba
\mathcal{L}_{SVV} &=&
\frac{g}{v}  \left( m_W W_\xi^- W^{\xi +}
+ \frac{m_Z}{2 c_{\theta_w}}\, Z_\xi Z^\xi \right)
\left( v_1 \rho_1 + v_2 \rho_2 \right)
\\ &=&
- g \left( m_W W_\xi^- W^{\xi +}
+ \frac{m_Z}{2 c_{\theta_w}}\, Z_\xi Z^\xi \right)
\left( h s_{\beta - \alpha} + H c_{\beta - \alpha} \right),
\ea
\es
where $\theta_w$ is Weinberg's angle.
Because of the LHC data we now know
that $\left| s_{\beta - \alpha} \right| \approx 1$.

The quark mass matrices are
\bs
\ba
M_n &=& \frac{v_1 e^{i \aleph_1} \Gamma_1 + v_2 e^{i \aleph_2} \Gamma_2}
{\sqrt{2}},
\\
M_p &=& \frac{v_1 e^{- i \aleph_1} \Delta_1 + v_2 e^{- i \aleph_2} \Delta_2}
{\sqrt{2}}.
\ea
\es
Let the unitary matrices $U_{L,R}^{n,p}$ bi-diagonalize $M_n$ and $M_p$ as
\bs
\label{uvifpgh}
\ba
   {U_L^n}^\dagger M_n U_R^n &=& M_d
   \equiv \mathrm{diag} \left( m_d,\ m_s,\ m_b \right),
\\
  {U_L^p}^\dagger M_p U_R^p &=& M_u
  \equiv \mathrm{diag} \left( m_u,\ m_c,\ m_t \right).
\ea
\es
The CKM matrix is
\be
V = {U_L^p}^\dagger U_L^n.
\ee
We define
\bs
\label{jdfofp1}
\ba
N_n &=& \frac{v_2 e^{i \aleph_1} \Gamma_1
  - v_1 e^{i \aleph_2} \Gamma_2}{\sqrt{2}},
\\
N_p &=& \frac{v_2 e^{- i \aleph_1} \Delta_1
  - v_1 e^{- i \aleph_2} \Delta_2}{\sqrt{2}},
\ea
\es
and
\bs
\label{jdfofp2}
\ba
N_d &=& {U_L^n}^\dagger N_n U_R^n,
\\
N_u &=& {U_L^p}^\dagger N_p U_R^p.
\ea
\es
Then,
the Yukawa interactions in the physical basis are given by
\bs
\label{jvwo}
\ba
\mathcal{L}_\mathrm{physical} &=&
\frac{iA}{v}\, \bar u \left( N_u P_R - N_u^\dagger P_L \right) u
\\ & &
+ \frac{iA}{v}\, \bar d \left( N_d^\dagger P_L - N_d P_R \right) d
\label{jsgifp} \\ & &
+ \frac{h}{v}\, \bar u \left[
  \left( s_{\beta - \alpha} M_u - c_{\beta - \alpha} N_u^\dagger \right) P_L
  + \left( s_{\beta - \alpha} M_u - c_{\beta - \alpha} N_u \right) P_R
  \right] u
  \label{eq:huu}
\\ & &
+ \frac{h}{v}\, \bar d \left[
  \left( s_{\beta - \alpha} M_d - c_{\beta - \alpha} N_d^\dagger \right) P_L
  + \left( s_{\beta - \alpha} M_d - c_{\beta - \alpha} N_d \right) P_R
  \right] d
\label{jsgifp2} \\ & &
+ \frac{H}{v}\, \bar u \left[
  \left( c_{\beta - \alpha} M_u + s_{\beta - \alpha} N_u^\dagger \right) P_L
  + \left( c_{\beta - \alpha} M_u + s_{\beta - \alpha} N_u \right) P_R
  \right] u
\\ & &
+ \frac{H}{v}\, \bar d \left[
  \left( c_{\beta - \alpha} M_d + s_{\beta - \alpha} N_d^\dagger \right) P_L
  + \left( c_{\beta - \alpha} M_d + s_{\beta - \alpha} N_d \right) P_R
  \right] d
\label{jsgifp3} \\ & &
+ \frac{\sqrt{2} H^+}{v}\,
\bar u \left( N_u^\dagger V P_L - V N_d P_R \right) d
\label{jhigfofdp} \\ & &
+ \frac{\sqrt{2} H^-}{v}\, \bar d \left( V^\dagger N_u P_R
- N_d^\dagger V^\dagger P_L \right) u,
\label{jhigfofdp2}
\ea
\es
where $P_L = \left. \left( 1 - \gamma_5 \right) \right/ 2$
and $P_R = \left. \left( 1 + \gamma_5 \right) \right/ 2$
are the projectors of chirality.
Also,
$u$ and $d$ are column vectors
subsuming the fields of the physical up-type and down-type quarks,
respectively.
In equation~\eqref{jvwo} we have omitted the Yukawa interactions
of the Goldstone bosons;
they have the same Lagrangian as in the SM.

In this paper we put forward a 2HDM
where $CP$ is conserved \emph{in the dimension-four terms}\/ of the Lagrangian,
hence $\theta_\mathrm{QCD} = 0$,
and $\det{\left( M_p M_n \right)}$ is real
both because of $CP$ and because of a $\mathbbm{Z}_3$ symmetry.
(Another symmetry-furnished 2HDM
that also purported to alleviate the strong $CP$ problem
was proposed long time ago~\cite{segre}.)
Remarkably,
our model provides for the absence of strong $CP$ violation
even at the one-loop level.

In section~\ref{sec2} we explain our model.
In section~\ref{sec4} we demonstrate that strong $CP$ violation vanishes
at the one-loop level in our model.
Section~\ref{sec:phen} is devoted to the phenomenological constraints
on the model.
Section~\ref{sec:conc} contains the main conclusions of our work.
Appendices~\ref{app:bsg},
\ref{app:mes},
and~\ref{app:Zbb} present some formulas
used in the analysis of section~\ref{sec:phen};
appendix~\ref{app:bench} gives a benchmark point
for the parameters of the model.

\section{The model} \label{sec2}

Our model is a 2HDM supplemented by the standard $CP$ symmetry
and by a $\mathbbm{Z}_3$ symmetry.
Let $\omega = \exp{\left( 2 i \pi / 3 \right)}$,
then the $\mathbbm{Z}_3$ symmetry reads
\bs
\label{z3}
\ba
& & \Phi_2 \to \omega^2 \Phi_2, \\
& & Q_{L1} \to \omega^2 Q_{L1}, \quad Q_{L2} \to \omega Q_{L2}, \\
& & n_{R3} \to \omega n_{R3}, \\
& & p_{R1} \to \omega p_{R1}, \quad p_{R2} \to \omega p_{R2}.
\label{pr}
\ea
\es
This represents just a slight change
from the $\mathbbm{Z}_3$ symmetry of the 2HDM of ref.~\cite{Ferreira:2011xc}.

Both $CP$ and $\mathbbm{Z}_3$ are softly broken by terms
in the quadratic part of the scalar potential. Soft breaking
  of a symmetry consists in that symmetry holding
  in all the Lagrangian terms of dimension higher than some value,
  but not holding for the Lagrangian terms of dimension
  smaller than,  or equal to, that value. In our case,
  both $CP$ and $\mathbbm{Z}_3$ hold for terms of dimension four
  but are broken by terms of dimension two,
  \textit{viz.}\ by the terms with coefficient $\mu_3$ in line~\eqref{guigo}.
  In principle,
  a model with a softly broken symmetry should eventually be justified
  through an ultraviolet completion,
  \textit{viz.}\ a more complete model,
  with extra fields active at higher energies,
  which effectively mimics at low energy scales
  the model with the softly-broken symmetry.
  Unfortunately,
  such a ultraviolet completion is often quite difficult
  to construct explicitly -- we attempted such a construction by adding
  singlet scalar fields to the theory, who would develop vevs at some
  high scale and then be ``integrated out", leaving the desired low
  energy potential with only two doublets. However, we were unable to
  build such extensions that left both the $CP$ and $\mathbbm{Z}_3$
  symmetries intact at low energies -- which of course does not mean
  such an UV completion does not exist.
  In the absence of any such explicit construction,
  a softly broken symmetry constitutes a strong,
  non-trivial assumption. This is,  certainly,
  a weakness of the model in this paper.

The softly broken $CP$ and $\mathbbm{Z}_3$ scalar potential is written as
\bs
\label{potent}
\ba
V &=&
\mu_1\, \Phi_1^\dagger \Phi_1
+ \mu_2\, \Phi_2^\dagger \Phi_2
- \mu_3 \left( e^{- i \aleph} \Phi_1^\dagger \Phi_2
+ e^{i \aleph} \Phi_2^\dagger \Phi_1 \right)
\label{guigo} \\ & &
+ \frac{\lambda_1}{2} \left( \Phi_1^\dagger \Phi_1 \right)^2
+ \frac{\lambda_2}{2} \left( \Phi_2^\dagger \Phi_2 \right)^2
+ \lambda_3\, \Phi_1^\dagger \Phi_1\, \Phi_2^\dagger \Phi_2
+ \lambda_4\, \Phi_1^\dagger \Phi_2\, \Phi_2^\dagger \Phi_1,
\ea
\es
where $\mu_3$ is real and positive by definition.
The terms with coefficient $\mu_3$ break the symmetry $\mathbbm{Z}_3$ softly.
The phase $\aleph$ breaks $CP$ softly.

The vacuum expectation values
$\left\langle 0 \left| \phi_a^0 \right| 0 \right\rangle
= v_a \exp{\left( i \aleph_a \right)} \left/ \sqrt{2} \right.$
have phases $\aleph_a$ such that $\aleph_2 - \aleph_1 = \aleph$
offsets the phase $- \aleph$ of the term
$- \mu_3 e^{- i \aleph} \Phi_1^\dagger \Phi_2$ of the scalar potential.
Thus,
there is one gauge-invariant vacuum phase that offsets
one phase in the potential,
with the consequence that the potential of the physical scalar fields
is $CP$-invariant,
in particular there is no mixing between the scalars $h$
and $H$ and the pseudoscalar $A$.
The stationarity equations for the vacuum are
\be
\mu_1 v_1^2 + \frac{\lambda_1}{2}\, v_1^4
= \mu_2 v_2^2 + \frac{\lambda_2}{2}\, v_2^4
= \mu_3 v_1 v_2 - \frac{\lambda_3 + \lambda_4}{2}\, v_1^2 v_2^2.
\ee
Referring to equation~\eqref{fmldpdi} and defining
\be
T_a = \frac{\rho_a^2 + \eta_a^2}{2} + \phi_a^- \phi_a^+
\ee
for $a = 1, 2$,
the potential is then
\bs
\label{fjidsosp}
\ba
V &=& - \frac{\lambda_1 v_1^4 + \lambda_2 v_2^4}{8}
- \frac{\lambda_3 + \lambda_4}{4}\, v_1^2 v_2^2
\\ & &
+ \mu_3 \left( \frac{v_2}{v_1}\, T_1 + \frac{v_1}{v_2}\, T_2
- \rho_1 \rho_2 - \eta_1 \eta_2 - \phi_1^- \phi_2^+ - \phi_2^- \phi_1^+ \right)
\\ & &
+ \sum_{a=1}^2 \frac{\lambda_a}{2} \left( v_a \rho_a + T_a \right)^2
+ \left( \lambda_3 + \lambda_4 \right) \left( v_1 \rho_1 + T_1 \right)
\left( v_2 \rho_2 + T_2 \right)
\\ & &
- \frac{\lambda_4}{2} \left\{
\left[ \left( v_1 + \rho_1 \right) \phi_2^-
  - \left( v_2 + \rho_2 \right) \phi_1^- \right]
\left[ \left( v_1 + \rho_1 \right) \phi_2^+
  - \left( v_2 + \rho_2 \right) \phi_1^+ \right]
\right. \\ & &
+ \left( \eta_1 \phi_2^- - \eta_2 \phi_1^- \right)
\left( \eta_1 \phi_2^+ - \eta_2 \phi_1^+ \right)
\\ & & \left.
+ i \left( \phi_2^- \phi_1^+ - \phi_1^- \phi_2^+ \right)
\left[ \eta_1 \left( v_2 + \rho_2 \right) - \eta_2 \left( v_1 + \rho_1 \right)
  \right]
\right\}.
\ea
\es
The potential~\eqref{fjidsosp} is invariant under the $CP$ transformation
\be
\label{cpcp}
CP: \quad \left\{ \begin{array}{rcl}
  \phi_a^- \left( x \right) &\to& \exp{\left( i \lambda \right)}\,
  \phi_a^+ \left( \bar x \right),
  \\
  \phi_a^+ \left( x \right) &\to& \exp{\left( - i \lambda \right)}\,
  \phi_a^- \left( \bar x \right),
  \\
  \rho_a \left( x \right) &\to& \rho_a \left( \bar x \right),
  \\
  \eta_a \left( x \right) &\to& - \eta_a \left( \bar x \right),
\end{array} \right.
\ee
for $a = 1, 2$,
where $x = \left( t,\ \vec r \right)$
and $\bar x = \left( t,\ - \vec r \right)$.
The phase $\lambda$ in the $CP$ transformation~\eqref{cpcp} is arbitrary.

The physical potential contains seven parameters $\mu_{1,2,3}$
and $\lambda_{1,2,3,4}$,
since the phase $\aleph$ in line~\eqref{guigo}
is cancelled out by the vacuum phase.
Instead of those seven parameters we will use as input $v = 246$\,GeV,
$m_h = 125$\,GeV,
the angles $\alpha$ and $\beta$,
and the masses $m_H$ of $H$,
$m_A$ of $A$,
and $m_{H^+}$ of $H^\pm$.
Then~\cite{Barroso:2012mj},
\bs
\label{lambalamba}
\ba
\mu_3 &=& m_A^2 s_\beta c_\beta,
\\
\lambda_1 &=& \frac{- m_A^2 s_\beta^2 + m_h^2 s_\alpha^2
  + m_H^2 c_\alpha^2}{v^2 c_\beta^2},
\\
\lambda_2 &=& \frac{- m_A^2 c_\beta^2 + m_h^2 c_\alpha^2
  + m_H^2 s_\alpha^2}{v^2 s_\beta^2},
\\
\lambda_3 &=& \frac{2 m_{H^+}^2 - m_A^2}{v^2}
+ \frac{\left( m_H^2 - m_h^2 \right) s_\alpha c_\alpha}{v^2 s_\beta c_\beta},
\label{eq:lamb}
\\
\lambda_4 &=& \frac{2 \left( m_A^2 - m_{H^+}^2 \right)}{v^2}.
\label{eq:lamb4}
\ea
\es

In order for the potential to be bounded from below,
one must impose the conditions~\cite{2HDM}
\be
\lambda_1 > 0, \quad
\lambda_2 > 0, \quad
\lambda_3 > - \sqrt{\lambda_1 \lambda_2}, \quad
\lambda_3 + \lambda_4 > - \sqrt{\lambda_1 \lambda_2}.
\label{eq:bfb}
\ee
In order to avoid
the situation of `panic vacuum'~\cite{Barroso:2012mj}
one must enforce the condition~\cite{Ivanov:2015nea}
\be
\frac{2 \mu_3}{v_1 v_2} > \lambda_3 + \lambda_4 - \sqrt{\lambda_1 \lambda_2}.
\label{condi2}
\ee
The conditions in order for tree-level unitarity not to be violated are
\bs
\label{unitunit}
\ba
\left| \lambda_1 \right| &<& 8 \pi, \label{eq:u1} \\
\left| \lambda_2 \right| &<& 8 \pi, \\
\left| \lambda_3 \right| &<& 8 \pi, \\
\left| \lambda_3  + \lambda_4 \right| &<& 8 \pi, \\
\left| \lambda_3  - \lambda_4 \right| &<& 8 \pi, \\
\left| \lambda_3  + 2 \lambda_4 \right| &<& 8 \pi, \\
\left| \lambda_1 + \lambda_2
+ \sqrt{\left( \lambda_1 - \lambda_2 \right)^2
  + 4 \lambda_4^2} \right| &<& 16 \pi, \\
\left| 3 \lambda_1 + 3 \lambda_2
+ \sqrt{9 \left( \lambda_1 - \lambda_2 \right)^2
  + 4 \left( 2 \lambda_3 + \lambda_4 \right)^2} \right| &<& 16 \pi.
\label{eq:un}
\ea
\es

Because of the $\mathbbm{Z}_3$ symmetry~\eqref{z3},
the matrices $\Gamma_a$ and $\Delta_a$ are
\bs
\label{cjviodp}
\ba
\Gamma_1 = \left( \begin{array}{ccc}
  0 & 0 & 0 \\ 0 & 0 & b_1 \\ d_1 & f_1 & 0 \end{array} \right),
& &
\Gamma_2 = \left( \begin{array}{ccc}
  d_2 & f_2 & 0 \\ 0 & 0 & 0 \\ 0 & 0 & b_2 \end{array} \right),
\label{eq:gam}
\\
\Delta_1 = \left( \begin{array}{ccc}
  0 & 0 & 0 \\ p_1 & q_1 & 0 \\ 0 & 0 & r_1\end{array} \right),
& &
\Delta_2 = \left( \begin{array}{ccc}
  p_2 & q_2 & 0 \\ 0 & 0 & r_2 \\ 0 & 0 & 0 \end{array} \right).
\label{eq:del}
\ea
\es
The dimensionless numbers $b_a$,
$d_a$,
$f_a$,
$p_a$,
$q_a$,
and $r_a$ ($a = 1, 2$) are real because of the $CP$ symmetry.
Clearly,
\bs
\ba
M_n &=&
\frac{1}{\sqrt{2}} \left( \begin{array}{ccc}
  d_2 v_2 e^{i \aleph_2} & f_2 v_2 e^{i \aleph_2} & 0 \\
  0 & 0 & b_1 v_1 e^{i \aleph_1} \\
  d_1 v_1 e^{i \aleph_1} & f_1 v_1 e^{i \aleph_1} & b_2 v_2 e^{i \aleph_2}
\end{array} \right),
\\*[1mm]
M_p &=& \frac{1}{\sqrt{2}} \left( \begin{array}{ccc}
  p_2 v_2 e^{- i \aleph_2} & q_2 v_2 e^{- i \aleph_2} & 0 \\
  p_1 v_1 e^{- i \aleph_1} & q_1 v_1 e^{- i \aleph_1} & r_2 v_2 e^{- i \aleph_2} \\
  0 & 0 & r_1 v_1 e^{- i \aleph_1}
\end{array} \right).
\ea
\es
Therefore
\be
\det{\left( M_n M_p \right)} =
\frac{v_1^4 v_2^2}{8}\, b_1 \left( d_1 f_2 - d_2 f_1 \right)
r_1 \left( p_2 q_1 - p_1 q_2 \right)
\ee
is real,
hence $\theta_\mathrm{QFD} = 0$.
Because of the assumed $CP$ invariance of the quartic part of the Lagrangian,
$\theta_\mathrm{QCD} = 0$ too.
Thus,
$\theta = \theta_\mathrm{QCD} + \theta_\mathrm{QFD} = 0$,
\textit{i.e.}\ there is no strong $CP$ violation at the tree level.

We now define
\bs
\label{phases1}
\ba
& & \beta_a \equiv \arg{b_a}, \quad
\delta_a \equiv \arg{d_a}, \quad
\varphi_a \equiv \arg{f_a},
\\
& & \pi_a \equiv \arg{p_a}, \quad
\chi_a \equiv \arg{q_a}, \quad
\varrho_a \equiv \arg{r_a},
\ea
\es
for $a = 1, 2$;
and
\bs
\label{phases2}
\ba
& & \Delta \beta \equiv \beta_2 - \beta_1, \quad
\Delta \delta \equiv \delta_2 - \delta_1, \quad
\Delta \varphi \equiv \varphi_2 - \varphi_1,
\\
& & \Delta \pi \equiv \pi_2 - \pi_1, \quad
\Delta \chi \equiv \chi_2 - \chi_1, \quad
\Delta \varrho \equiv \varrho_2 - \varrho_1.
\ea
\es
All the phases in equations~\eqref{phases1} and~\eqref{phases2}
are either $0$ or $\pi$ because $b_a, d_a, \ldots, r_a$ are real.
We define the diagonal matrices
\bs
\ba
X_{Ln} &=& \mathrm{diag} \left( 1,\
e^{i \left( 2 \aleph + \Delta \delta + \Delta \beta \right)},\
e^{i \left( \aleph + \Delta \delta \right)}
\right),
\\
X_{Rn} &=& \mathrm{diag} \left(
e^{i \left( - \aleph_2 - \delta_2 \right)},\
e^{i \left( - \aleph_2 - \varphi_2 \right)},\
e^{i \left( \aleph_1 - 2 \aleph_2 - \Delta \delta - \beta_2 \right)}
\right),
\\
X_{Lp} &=& \mathrm{diag} \left( 1,\
e^{i \left( - \aleph + \Delta \pi \right)},\
e^{i \left( - 2 \aleph + \Delta \pi + \Delta \varrho \right)}
\right),
\\
X_{Rp} &=& \mathrm{diag} \left(
e^{i \left( \aleph_2 - \pi_2 \right)},\
e^{i \left( \aleph_2 - \chi_2 \right)},\
e^{i \left( 2 \aleph_2 - \aleph_1 - \Delta \pi - \varrho_2 \right)}
\right),
\ea
\es
where $\aleph = \aleph_2 - \aleph_1$.
We then have
\bs
\label{primes}
\ba
X_{Ln} M_n X_{Rn} &=&
\frac{1}{\sqrt{2}} \left( \begin{array}{ccc}
  \left| d_2 v_2 \right| & \left| f_2 v_2 \right| & 0 \\
  0 & 0 & \left| b_1 v_1 \right| \\
  \left| d_1 v_1 \right| & \left| f_1 v_1 \right|
  e^{i \left( \Delta \delta - \Delta \varphi \right)} & \left| b_2 v_2 \right|
\end{array} \right)
\\ &\equiv& M_n^\prime,
\\*[1mm]
X_{Lp} M_p X_{Rp} &=&
\frac{1}{\sqrt{2}} \left( \begin{array}{ccc}
  \left| p_2 v_2 \right| & \left| q_2 v_2 \right| & 0 \\
  \left| p_1 v_1 \right| & \left| q_1 v_1 \right|
  e^{i \left( \Delta \pi - \Delta \chi \right)} & \left| r_2 v_2 \right| \\
  0 & 0 & \left| r_1 v_1 \right|
\end{array} \right)
\\ &\equiv& M_p^\prime.
\ea
\es
The matrices $M_n^\prime$ and $M_p^\prime$ are real,
therefore they may be bi-diagonalized through real orthogonal matrices $O_{Ln}$,
$O_{Rn}$,
$O_{Lp}$,
and $O_{Rp}$ as
\be
\label{lnlp}
O_{Ln} M_n^\prime O_{Rn} = M_d, \quad O_{Lp} M_p^\prime O_{Rp} = M_u.
\ee
Therefore,
in the notation of equations~\eqref{uvifpgh},
\be
{U_L^n}^\dagger = O_{Ln} X_{Ln}, \quad
U_R^n = X_{Rn} O_{Rn}, \quad
{U_L^p}^\dagger = O_{Lp} X_{Lp} \quad
U_R^p = X_{Rp} O_{Rp}.
\ee
The CKM matrix is then
\be
\label{djkdpfd}
V = O_{Lp} \times \mathrm{diag} \left( 1,\
e^{i \left( - 3 \aleph + \Delta \pi - \Delta \delta - \Delta \beta \right)},\
e^{i \left( - 3 \aleph + \Delta \pi - \Delta \delta + \Delta \varrho \right)} \right)
\times O_{Ln}^T.
\ee
One sees that the CKM matrix is \emph{complex}\/
because of the presence of the phase $3 \aleph$.

When we compute the matrices $N_p$ and $N_n$
defined in equations~\eqref{jdfofp1},
we find that
\bs
\label{primes2}
\ba
X_{Ln} N_n X_{Rn} &=& \frac{1}{\sqrt{2}} \left( \begin{array}{ccc}
  - \left| d_2 v_1 \right| & - \left| f_2 v_1 \right| & 0 \\
  0 & 0 & \left| b_1 v_2 \right| \\
  \left| d_1 v_2 \right| & \left| f_1 v_2 \right|
  e^{i \left( \Delta \delta - \Delta \varphi \right)} & - \left| b_2 v_1 \right|
\end{array} \right)
\\ &\equiv& N_n^\prime,
\\*[1mm]
X_{Lp} N_p X_{Rp} &=& \frac{1}{\sqrt{2}} \left( \begin{array}{ccc}
  - \left| p_2 v_1 \right| & - \left| q_2 v_1 \right| & 0 \\
  \left| p_1 v_2 \right| & \left| q_1 v_2 \right|
  e^{i \left( \Delta \pi - \Delta \chi \right)} & - \left| r_2 v_1 \right| \\
  0 & 0 & \left| r_1 v_2 \right|
\end{array} \right)
\\ &\equiv& N_p^\prime,
\ea
\es
and then,
from equations~\eqref{jdfofp2},
\be
\label{n4}
N_d = O_{Ln} N_n^\prime O_{Rn},
\quad
N_u = O_{Lp} N_p^\prime O_{Rp}.
\ee
The matrices $N_d$ and $N_u$ are \emph{real}.

Thus,
in our model
\begin{enumerate}
\item The CKM matrix is complex.
\item The matrices $N_u$ and $N_d$ are real.
\item There is one pseudoscalar $A$ that does not mix with the scalars
  $h$ and $H$.
\item There is no $CP$ violation in the cubic and quartic interactions
  of the scalars.
\end{enumerate}
In our model $CP$ violation is located solely in the CKM matrix
and originates entirely in the phase $3 \aleph$.
This is the same that happened in the model of ref.~\cite{Ferreira:2011xc};
however,
in that model there was strong $CP$ violation,
while in the present model strong $CP$ violation is absent at the tree level.

\section{No strong $CP$ violation at the one-loop level} \label{sec4}

At one-loop level the diagonal and real quark mass matrices $M_q$
(where $q$ may be either $u$ or $d$)
get corrected by self energy diagrams:
$M_q \to M_q + \Sigma_q$.
If the diagonal elements of $\Sigma_q$ are complex,
then $\mathrm{Im} \left[ \mathrm{tr} \left( M_q^{-1} \Sigma_q \right) \right]$
may be nonzero and strong $CP$ violation may arise.

In our model there are no complex phases except in the CKM matrix.
Since the matrices $N_q$ are real,
and since the scalars $h$ and $H$ do not mix with the pseudoscalar $A$,
the $\Sigma_q$ generated through the emission and reabsorption (E\&R)
by the quarks of either $h$ or $H$ or $A$  are real,
hence innocuous.
The same happens with the $\Sigma_q$ generated through the E\&R
of $Z^0$ gauge bosons.
On the other hand,
diagrams with the E\&R of $W^\pm$ gauge bosons
do not generate mass renormalization
(they just produce wavefunction renormalization),
since the coupling of $W^\pm$ to the quarks is purely left-handed.
Therefore,
the only diagrams where the complex matrix $V$ arises,
and might produce complex $\Sigma_q$,
are the ones with E\&R of charged scalars $H^\pm$.

The Yukawa interactions of the charged scalars are given by
lines~\eqref{jhigfofdp} and~\eqref{jhigfofdp2}.
They contain two complex matrices,
$X \equiv N_u^\dagger V$ and $Y \equiv V N_d$.
The one-loop self-energy of an up-type quark $u_\alpha$
caused by the E\&R of $H^+$ and a down-type quark $d_j$ is
\bs
\ba
- i \Sigma_\alpha \left( p \right) &=&
\frac{2}{v^2}\, \mu^{4-d}
\int \! \frac{\mathrm{d}^d k}{\left( 2 \pi \right)^d}\,
\frac{1}{k^2 - m_j^2}\, \frac{1}{\left( k - p \right)^2 - m_{H^+}^2}
\\ & & \times
\left( X_{\alpha j} P_L - Y_{\alpha j} P_R \right)
\left( \not \! k + m_j \right)
\left( X^\dagger_{j \alpha} P_R - Y^\dagger_{j \alpha} P_L \right)
\\ &=&
\frac{2}{v^2}\, \mu^{4-d}
\int \! \frac{\mathrm{d}^d k}{\left( 2 \pi \right)^d}
\int_0^1 \mathrm{d}x\, \frac{1}{\left( k^2 - \Delta_j \right)^2}
\\ & & \times
\left(
x \left| X_{\alpha j} \right|^2 \not \! p P_R
+ x \left| Y_{\alpha j} \right|^2 \not \! p P_L
- X_{\alpha j} m_j Y^\dagger_{j \alpha} P_L
- Y_{\alpha j} m_j X^\dagger_{j \alpha} P_R
\right), \hspace*{7mm}
\ea
\es
where
\be
\Delta_j = p^2 x^2 + \left( - p^2 + m_{H^+}^2 - m_j^2 \right) x + m_j^2
\ee
and we perform the computation in a space--time of dimension $d$.
Thus,
the only potentially complex part of the self-energy is
\be
\label{mldps}
\Sigma_\alpha \left( \not \! p \to 0 \right) =
- \int_0^1 \mathrm{d} x\
\frac{X_{\alpha j} m_j Y^\dagger_{j \alpha} P_L
  + Y_{\alpha j} m_j X^\dagger_{j \alpha} P_R}{8 \pi^2 v^2}
\left( \frac{2}{4 - d} - \gamma - \ln{\frac{\Delta_j}{4 \pi \mu^2}} \right),
\ee
where $\gamma$ is Euler--Mascheroni's constant.
One must sum the expression in the right-hand side of equation~\eqref{mldps}
over the flavour $j$ of the quark $d_j$.

The one-loop value of the strong-$CP$ parameter $\theta$
is $\mathrm{Im} \left[ \mathrm{tr} \left( M_u^{-1} \Sigma_u
  + M_d^{-1} \Sigma_d \right) \right]$.
The diagonal matrix elements of $M_u^{-1} \Sigma_u$
are $m_\alpha^{-1}\, \Sigma_\alpha \left( \not \! p \to 0 \right)$.
Now,
\bs
\ba
\sum_{\alpha} \frac{X_{\alpha i} Y^\dagger_{j \alpha}}{m_\alpha}
&=&
\sum_{\alpha} \left( N_d^\dagger V^\dagger \right)_{j \alpha}\, \frac{1}{m_\alpha}\,
\left( N_u^\dagger V \right)_{\alpha i}
\\ &=&
\left( N_d^\dagger V^\dagger {M_u^{-1}}^\dagger N_u^\dagger V \right)_{ji}
\\ &=&
\left( {U_R^n}^\dagger N_n^\dagger {M_p^{-1}}^\dagger N_p^\dagger U_L^n \right)_{ji}.
\ea
\es
Therefore,
\bs
\ba
\sum_{\alpha, j} \frac{X_{\alpha j} m_j Y^\dagger_{j \alpha}\,
  f \left( m_j^2 \right)}{m_\alpha}
&=&
\sum_j f \left( m_j^2 \right) m_j
\left( {U_R^n}^\dagger N_n^\dagger {M_p^{-1}}^\dagger N_p^\dagger U_L^n \right)_{jj}
\\ &=&
\mathrm{tr} \left[ U_L^n\, f \left( M_d M_d^\dagger \right) M_d {U_R^n}^\dagger
N_n^\dagger {M_p^{-1}}^\dagger N_p^\dagger \right]
\\ &=&
\mathrm{tr} \left[ f \left( M_n M_n^\dagger \right)
M_n N_n^\dagger {M_p^{-1}}^\dagger N_p^\dagger \right].
\ea
\es
One easily finds that,
in our model,
both matrices $M_n M_n^\dagger$ and $M_n N_n^\dagger {M_p^{-1}}^\dagger N_p^\dagger$,
and all of their products too,
have a structure of phases of the form
\be
\left( \begin{array}{ccc}
  0 & 2 \aleph & \aleph \\
  - 2 \aleph & 0 & - \aleph \\
  - \aleph & \aleph & 0
\end{array} \right),
\ee
where $\aleph = \aleph_2 - \aleph_1$.
Therefore,
the diagonal matrix elements of $f \left( M_n M_n^\dagger \right)
M_n N_n^\dagger {M_p^{-1}}^\dagger N_p^\dagger$,
and hence its trace,
are real,
no matter what the function $f$ is.

In this way we have demonstrated that
$\mathrm{tr} \left( M_u^{-1} \Sigma_u \right)$ is real.
In a similar way one may show that
$\mathrm{tr} \left( M_d^{-1} \Sigma_d \right)$ is also real,
hence strong $CP$ violation vanishes at the one-loop level in our model.

\section{Phenomenological analysis of the model}
\label{sec:phen}

\subsection{Constraints}
\label{sec:phen0}

We proceed to analyse how our model conforms to the experimental results.
The model has tree-level flavour-changing neutral currents (FCNC)
coupling to the scalars,
so there is a wealth of flavour-physics observables
that need to be taken into account whilst performing a fit of the model
to the experimental data.
Our procedure involves a global fit of the model's parameters,
simultaneously requiring compliance
with the theoretical and experimental bounds from the gauge,
scalar,
and fermionic sectors.

One may rotate the right-handed quarks $n_{R1}$ and $n_{R2}$ between themselves
in such a way that the entry $f_2$ of the Yukawa-coupling matrix $\Gamma_2$
becomes zero.
Similarly,
one may rotate $p_{R1}$ and $p_{R2}$ so that $q_2$
becomes zero.\footnote{Notice that $n_{R1}$ and $n_{R2}$
  transform in the same way under the $\mathbbm{Z}_3$ symmetry~\eqref{z3},
  and $p_{R1}$ and $p_{R2}$ also transform in the same way under
  that symmetry.}$^,$\footnote{With $f_2 = q_2 = 0$,
  the phases $\Delta \varphi$ and $\Delta \chi$ in equations~\eqref{phases2}
  become meaningless.
  That has no impact on our reasonings,
  in particular the matrix $V$ in equation~\eqref{djkdpfd}
  does not depend on those phases.}
We use as input the ten entries
$b_1$,
$d_1$,
$f_1$,
$b_2$,
$d_2$,
$p_1$,
$q_1$,
$r_1$,
$p_2$,
and $r_2$ ($f_2$ and $q_2$ are set to zero)
of the Yukawa-coupling matrices~\eqref{cjviodp},
allowing those entries to be either positive or negative.
We further input the $CP$-violating phase $\aleph$.
We fit these eleven parameters in order to reproduce
the quark masses~\cite{Tanabashi:2018oca}\footnote{We have doubled
  the uncertainty intervals quoted in ref.~\cite{Tanabashi:2018oca}
  for the masses of the light quarks $u$,
  $d$,
  and $s$;
  we have done this because of the large theoretical indefinition,
  due to QCD considerations,
  as to what exactly should be interpreted as the value of those masses.}
\bs
\label{quarkmasses}
\ba
m_u &=& \left( 2.2 \pm 2 \times 0.6 \right) \mathrm{MeV}, \\
m_d &=& \left( 4.7 \pm 2 \times 0.5 \right) \mathrm{MeV}, \\
m_s &=& \left( 96 \pm 2 \times 8 \right) \mathrm{MeV}, \\
m_c &=& \left( 1.28 \pm 0.03 \right) \mathrm{GeV}, \\
m_b &=& \left( 4.18 \pm 0.04 \right) \mathrm{GeV}, \\
m_t &=& \left( 173.2 \pm 0.6 \right) \mathrm{GeV},
\ea
\es
and the CKM-matrix observables~\cite{Tanabashi:2018oca}
\bs
\label{ckmmatrix}
\ba
\left| V_{us} \right| &=& 0.2243 \pm 0.0005, \\
\left| V_{cb} \right| &=& 0.0422 \pm 0.0008, \\
\left| V_{ub} \right| &=& 0.00394 \pm 0.00036, \\
\gamma \equiv \arg{\left( - \frac{V_{ud} V^\ast_{ub}}{V_{cd} V^\ast_{cb}} \right)}
&=& \left( 73.5 \pm 5.5 \right)^\circ.
\ea
\es
We furthermore input $\tan{\beta} = v_2 / v_1$.
We compute the matrices $N_n^\prime$ and $N_p^\prime$
through equations~\eqref{primes2},
and the matrices $N_d$ and $N_u$ through equations~\eqref{n4}.
Finally,
we input $\alpha$ and get to know $\mathcal{L}_\mathrm{physical}$
in equation~\eqref{jvwo}.\footnote{The angle $\alpha$
may be restricted to lie either in the first quadrant
or in the fourth quadrant~\cite{2HDM}.}

In the scalar potential~\eqref{potent}
there are seven independent parameters $\mu_{1,2,3}$ and $\lambda_{1,2,3,4}$.
We input instead the seven observables
\begin{enumerate}
\item $v = \sqrt{v_1^2 + v_2^2} = 246$\,GeV,
  which produces the correct masses for the electroweak gauge bosons $W^\pm$
  and $Z^0$;
\item the lightest $CP$-even-scalar mass $m_h = 125$\,GeV,
corresponding to the Higgs boson observed at the LHC;
\item the angle $\beta = \arctan{\left( v_2 / v_1 \right)}$;
\item the angle $\alpha$;
\item the remaining scalar masses---$m_H$ of the second CP-even scalar,
$m_A$ of the pseudoscalar,
and $m_{H^+}$ of the charged scalar.
\end{enumerate}
The last five parameters must be found through the fitting procedure.
We have constrained $m_{H^+}$ to be larger than 100\,GeV
and $m_H$ and $m_A$ to be larger than 130\,GeV.
We have furthermore assumed all three masses to be smaller than 1.2\,TeV;
values of the masses larger than 1.2\,TeV
would certainly be allowed by the fitting procedure.

The quartic couplings of the model
are determined via equations~\eqref{lambalamba}.
We check that the scalar potential is bounded from below,
that it does not have a panic vacuum,
and that it satisfies unitarity,
\textit{viz.}\ we check conditions~(\ref{eq:bfb}, \ref{condi2}, \ref{unitunit}).
The constraints from the electroweak oblique parameters $S$ and $T$
are also imposed,
by using the expressions for the 2HDM
in refs.~\cite{Toussaint:1978zm,Kanemura:2011sj}.

We will now go into detail about the further constraints that we have imposed.
\begin{itemize}
\item We implement the $b \to s \gamma$ bound
  described in appendix~\ref{app:bsg},
  including the contributions from both the neutral and the charged scalars.
\item The most relevant bounds
  on the off-diagonal entries of the matrices $N_d$ and $N_u$
  come from flavour-physics observables,
  specifically the $K$,
  $B_d$,
  $B_s$,
  and $D$ neutral-meson mass differences,
  and the $CP$-violating parameter $\epsilon_K$.
  We detail the computation of those quantities,
  and the requirements on them that we use in our fit,
  in appendix~\ref{app:mes}.
\item The $Z \to b \bar{b}$ constraints described in appendix~\ref{app:Zbb}
  are also taken into account.
  In this case we use only the charged-scalar contributions;
  the neutral-scalar ones should be negligible.
\item For the regions of parameter space where $m_t > m_q + m_{H^+}$,
  $q$ being a down-type quark,
  or where $m_t > m_q + m_S$,
  $q$ being either $c$ or $u$ and $S$ being either $h$ or $H$ or $A$,
  we require that the branching ratio for each of the kinematically viable
  $t \to $ \textit{light quark}\,$+$\,\textit{scalar} decays
  be smaller than $5\times 10^{-3}$,
  in accordance with the current results on FCNC top decays
  and on the total top-quark width~\cite{Tanabashi:2018oca}.
\item In order that the scalar $h$ of our model
  complies with the observational data
  from the LHC---it should be SM-like in its behaviour---we require
  that its couplings to the electroweak gauge bosons
  and to the top and bottom quarks do not deviate significantly
  from the SM expectations.
  We achieve this by focusing on the coupling modifiers $\kappa_X$ defined as
  $g_{hZZ} = \kappa_Z\, g_{hZZ}^\mathrm{SM}$,
  $g_{hWW} = \kappa_W\, g_{hWW}^\mathrm{SM}$,
  $g_{ht\bar{t}} = \kappa_t\, g_{ht\bar{t}}^\mathrm{SM}$,
  and $g_{hb\bar{b}} = \kappa_b\, g_{hb\bar{b}}^\mathrm{SM}$.
  In our model $\kappa_Z = \kappa_W \equiv \kappa_V$ and
  \be
 \kappa_V = s_{\beta - \alpha} , \quad
  \kappa_t = s_{\beta - \alpha}
  - \frac{c_{\beta - \alpha} \left( N_u \right)_{33}}{m_t}, \quad
  \kappa_b = s_{\beta - \alpha}
  - \frac{c_{\beta - \alpha} \left( N_d \right)_{33}}{m_b},
  \label{eq:ktkb}
  \ee
  \textit{cf.}\ equations~\eqref{eq:huu} and~\eqref{jsgifp2}.
  In the first stage of the fit we constrain
  these couplings to obey $0.8 \leq \kappa_V \leq 1$,
  $0.8 \leq \kappa_t \leq 1.2$,
  and $0.8 \leq \left| \kappa_b \right| \leq 1.2$,\footnote{Notice that
    we allow the possibility of a
    `wrong-sign regime'~\cite{Ferreira:2014naa,Ferreira:2014dya}
    in the bottom-quark coupling.
    However,
    the combination of cuts applied to the model
    ends up not allowing for that regime.}$^,$\footnote{Notice that
    $\kappa_V$ and $\kappa_t$ must have the same sign,
    otherwise one would expect a huge variation in the $h$ diphoton width,
    in disagreement with the SM-like observed values.}
  in order to roughly reproduce the LHC results.
  A second stage of the analysis further constrains these couplings,
  as detailed below.
\end{itemize}

A numerical scan of the parameter space of the model,
in both the scalar and Yukawa sectors,
was performed
to discover points that obey all the constraints described above.
It must be stressed that we have introduced nowhere in our scan
a `no-fine-tuning' assumption:
we have tolerated any set of input values that led to the right outputs,
even if either the input values or any intermediate computations
displayed either `fine-tunings' or `unnatural cancelations'.
Strong fine-tunings are often required
in order to fit the $D$-meson mass difference constraint
whenever $m_H$ and $m_A$ are not very high;
for the other constraints,
fine-tunings are at most moderate
and do not occur at all for many points in our fit.
As a matter of fact,
even for the $D$-meson mass difference,
there are many choices of parameters for which no fine-tuning is necessary
and one of the scalars has relatively low mass;
one such case is presented in appendix~D.

With those points we have proceeded
to compute the LHC production cross sections
of the neutral scalars in the model,
using the software {\tt SusHI}~\cite{Harlander:2012pb,Harlander:2016hcx}
to include the NNLO QCD corrections.
We have limited ourselves to the gluon--gluon production process,
which is the dominant one in the LHC environment.
Regarding the vector boson-fusion process,
no differences will occur in this model \textit{vis a vis}\ the usual 2HDM,
as the couplings of the scalars to the gauge bosons
are the same in both models.

The results for $h$ are expressed in terms of the ratios
\be
\mu_X = \frac{\sigma \left( pp \to h \right)\,
\text{BR} \left( h \to X \right)}{\sigma^\mathrm{SM} \left( pp \to h \right)\,
\text{BR}^\mathrm{SM} \left( h \to X \right)},
\label{eq:mux}
\ee
where $X$ may be either $Z^0 Z^0$,
$W^+ W^-$,
$b \bar b$,
$\tau \bar \tau$,
or $\gamma \gamma$.
The value $\mu_X = 1$ indicates exact SM-like behaviour.
We require that all the $\mu_X$ be within 20\% of 1,
which is a fair description of the current LHC results,
taking into account the uncertainties.
With this imposition,
the ranges of variation of $\kappa_V  = s_{\beta - \alpha}$,
$\kappa_t$,
and $\kappa_b$ become much smaller than initially allowed in the fit:
we obtain $0.929 \leq \kappa_V \leq 1$ and
$0.952 \leq \left\{ \kappa_t, \kappa_b \right\} \leq 1.04$.
For comparison,
we will also present results for the tighter constraint
$\left| \mu_X - 1 \right| < 0.1$.

In principle,
we should also consider the leptons.
The $\mathbbm{Z}_3$ symmetry in the quark sector
must be extended to the leptonic sector.
Since flavour violation with leptons is much more constrained than with quarks,
the best choice would be to extend $\mathbbm{Z}_3$ to the leptonic sector
in a way identical to the flavour-preserving 2HDMs,
allowing only one of the two doublets $\Phi_a$
to couple to the leptons and give them mass.
We would then have have two possibilities
for the couplings of the scalars to the charged leptons---either
$\Phi_1$ couples to the charged leptons or $\Phi_2$ does.
The coupling modifier $\kappa_\tau = g_{h\tau\bar{\tau}}
\left/ g_{h\tau\bar{\tau}}^\mathrm{SM} \right.$ is given by
\be
\kappa_\tau = \frac{\cos{\alpha}}{\sin{\beta}}
\quad \mbox{and} \quad
\kappa_\tau = - \frac{\sin{\alpha}}{\cos{\beta}}
\label{eq:ktau}
\ee
for the first and second choices,
respectively.
For definiteness,
in our fit we have adopted the second option in equation~\eqref{eq:ktau},
\textit{viz.}\/ we have imposed
\be
0.8 < \left| \frac{\sin{\alpha}}{\cos{\beta}} \right| < 1.2
\ee
to the points in our fit. However,
since the extension of our model to the leptonic sector
is largely arbitrary,
we have refrained from taking into account
any other constraints on our model
that might arise from processes involving leptons.
We point out, though,
that flavour-changing constraints from processes
like $K_L \to \mu^+ \mu^-$
or $B_s \to \mu^+ \mu^-$ may pose serious challenges to our model.

\subsection{General results}

The bounds on the scalar sector---unitarity,
oblique parameters,
and vacuum stability---produce the same contraints on the model's parameters
than those found in the usual version of the 2HDM.
On the other hand,
since the symmetry that we are considering
affects in a non-trivial way the quark Yukawa matrices,
there are major differences relative to other 2HDMs
when the flavour-physics bounds are imposed.
The flavour constraints from meson observables
and from the top-quark FCNC decays and total width,
previously described,
constrain severely the magnitudes of the off-diagonal elements
of the matrices $N_u$ and $N_d$.
Our fit achieves to keep FCNC under control
even with extra scalars of {\em ``low''}\/ masses---$H$ and $A$
may have masses below 500\,GeV.
This is in contrast with the often-made assumption
that models with tree-level FCNC imply masses above 1\,TeV;
this had already been shown not to necessarily apply
in the previous version of the current model~\cite{Ferreira:2011xc}.
In ref.~\cite{Nebot:2015wsa}
it has been argued that contributions from the scalar and pseudoscalar particles
($H$ and $A$,
respectively)
to FCNC meson observables tend to cancel each other;
we have explicitly observed that,
for many points in our fit
(\textit{cf.}\ the point given in appendix~\ref{app:bench}),
the arguments of ref.~\cite{Nebot:2015wsa} apply,
and this is the reason why masses of the extra scalars
lower than 1\,TeV are possible.

One consequence of the present model is the fact that the scalar $h$,
which we have taken to be the 125\,GeV state observed at the LHC,
has tree-level FCNCs,
as indicated by its Yukawa interactions in lines~\eqref{eq:huu}
and~\eqref{jsgifp2}.
Thus,
unlike in the (tree-level) SM,
$h$ has the possibility of FCNC decays to final states $s \bar b$,
$d \bar b$,
$d \bar s$,
$u \bar c$,
and their charge-conjugate states.
However,
for all the points resulting from our fit,
these decays are extremely suppressed---the sum
of the branching ratios for all of them
being at most $2 \times 10^{-6}$ but usually much lower.
Therefore,
in our model the FCNC decays of the 125\,GeV-particle
are impossible to observe at the LHC,
and almost certainly even at future $e^+ e^-$ colliders such as the ILC;
the existence of those decays has no measurable impact
on the phenomenology of the scalar $h$.
The FCNC also raise the possibility of alternative production mechanisms
for $h$,
such as $d \bar s \to h$ or $u \bar c \to h$;
such production channels would be favoured by larger proton PDFs
relatively to the SM production mechanism $b \bar b \to h$.
However,
once again these FCNC processes are found to be extremely small in our fit.
According to~\eqref{eq:huu} and~\eqref{jsgifp2},
all the FCNC $h$ interactions are suppressed
by their proportionality to $c_{\beta - \alpha}$,
which is required to be quite small
by the SM-like behaviour of the 125\,GeV scalar $h$,
by the ratio between a light-quark mass and $v = 246$\,GeV,
and,
sometimes,
by the smallish off-diagonal $N_d$ and $N_u$ matrix elements
induced by compliance with meson-physics bounds.

\begin{figure}
\begin{center}
\includegraphics[height=8cm,angle=0]{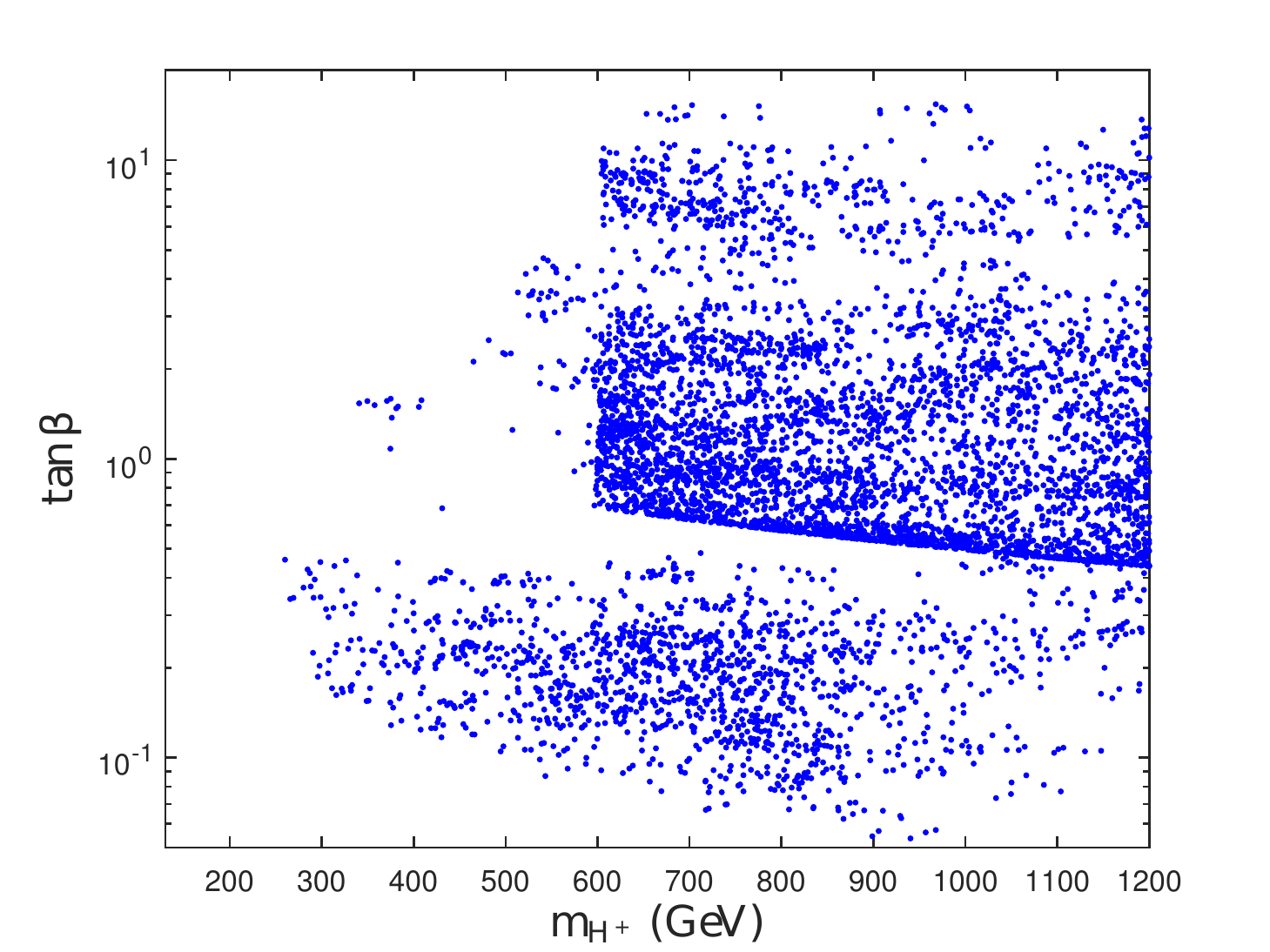}
\end{center}
\vspace{-5mm}
\caption{$v_2 / v_1$ \textit{versus}\/ the charged-Higgs mass
for the parameter-space points
that survived all the theoretical and experimental constraints.}
\label{fig:tanbmch}
\end{figure}

In fig.~\ref{fig:tanbmch} we show the points generated by our fit,
displayed on the $\tan{\beta}$--$m_{H^+}$ plane.
(The observed low density of points
is merely a consequence of the difficulty
in achieving good fits---further searches would yield more points
and fill many more regions in the plot;
lack of points in some areas has no physical meaning,
it is just an artifact of the limited parameter space scan.)
A clear conclusion from fig.~\ref{fig:tanbmch}
is that $1/20 < \tan{\beta} < 20$ in our model\footnote{As is plain
  in equations~\eqref{cjviodp},
  in our model there is a symmetry between the Yukawa couplings of $\Phi_1$
  and the ones of $\Phi_2$,
  so that,
  for any given $t$,
  $\tan{\beta} = t$ is just as (im)possible to achieve as $\tan{\beta} = 1/t$.
  This is in contrast to what happens in the usual 2HDMs types~I and~II.};
this is mainly a consequence of the $b \to s \gamma$ bounds and,
to a lesser degree,
of the $Z \to b \bar b$ bounds.
One also sees in fig.~\ref{fig:tanbmch} that
values of the charged-scalar mass as low as 130\,GeV are easily attained;
this is in stark contrast with the findings for the type~II 2HDM,
where a lower bound on $m_{H^+}$ of roughly 580\,GeV exists.
Unlike in the usual type-I and type-II 2HDMs,
in our model the quark mass matrices
do not emerge from the Yukawa couplings to a single scalar doublet,
but rather from the couplings to both $\Phi_1$ and $\Phi_2$.
As such,
although we employ the standard definition $\tan{\beta} = v_2/v_1$,
the usual wisdom about the values of this parameter does not apply.

\begin{figure}[t]
\begin{tabular}{cc}
\includegraphics[height=6cm,angle=0]{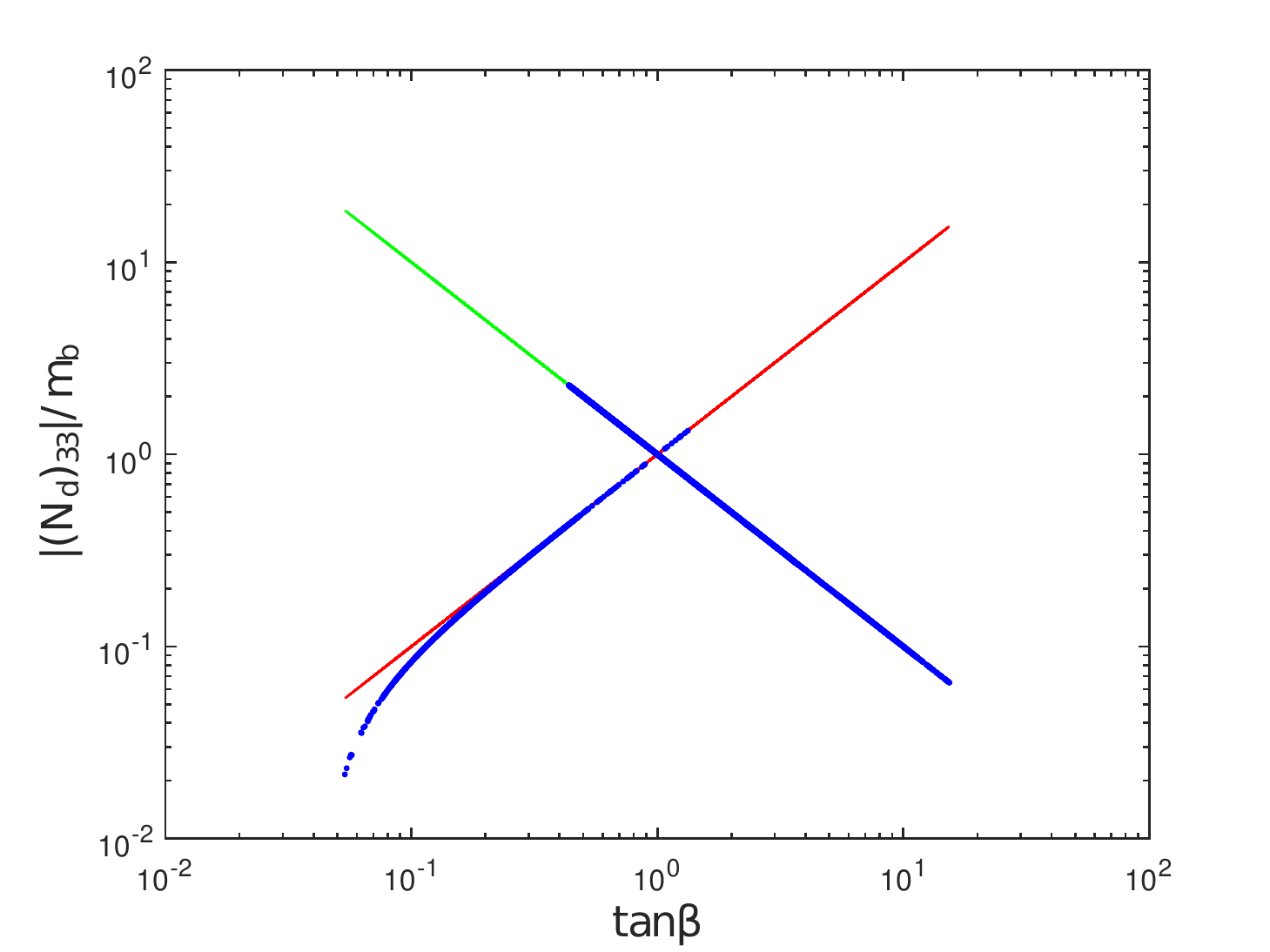} &
\includegraphics[height=6cm,angle=0]{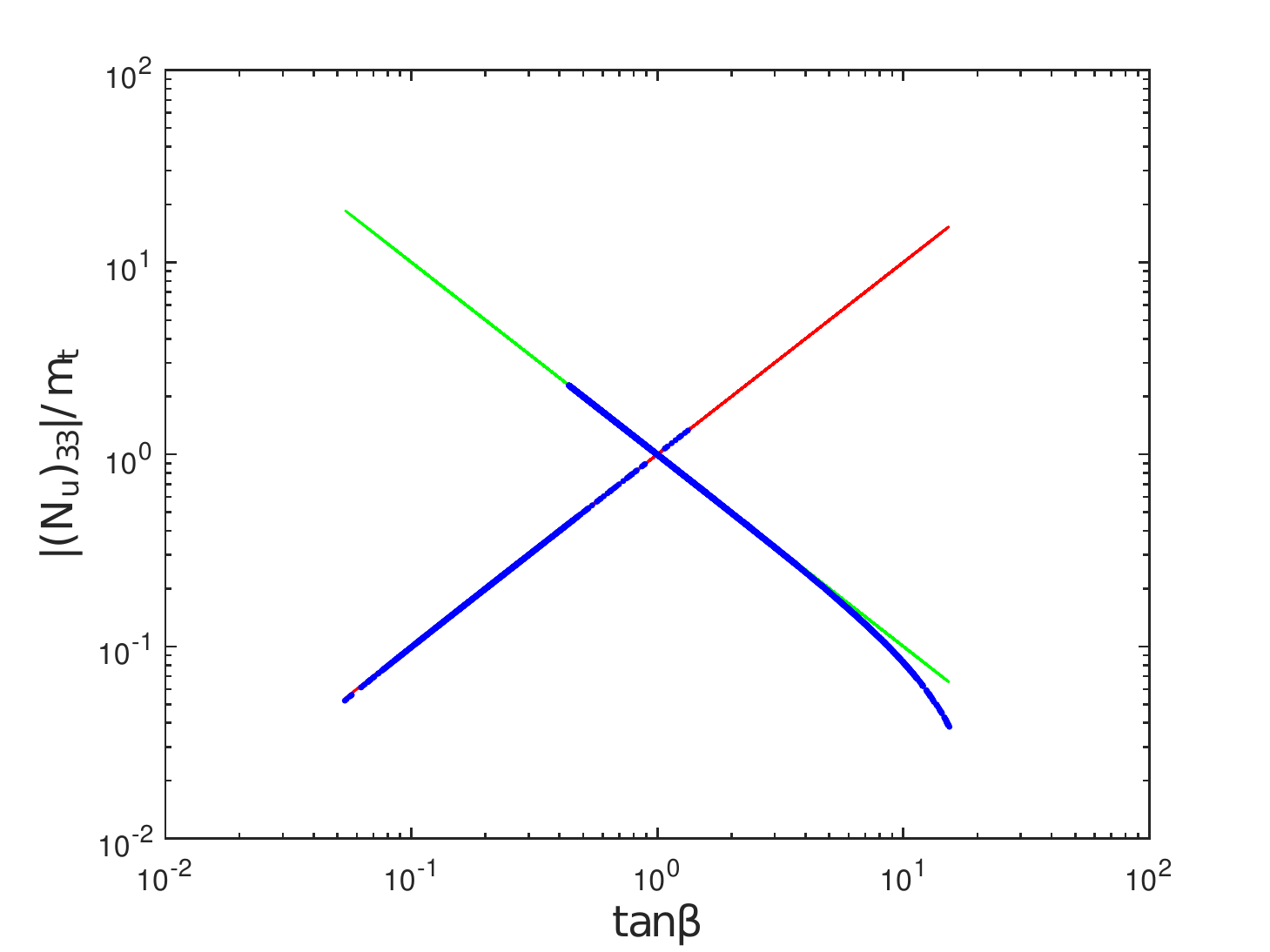} \\
(a) & (b)
\end{tabular}
\caption{(a) $\left. \left| \left( N_d \right)_{33} \right| \right/ m_b$
  and (b) $\left. \left| \left( N_u \right)_{33} \right| \right/ m_t$
  \textit{versus} $\tan{\beta}$ for the parameter space points (in blue)
  which survive all the theoretical and experimental constraints.
  The green lines correspond to $y = \cot{\beta}$
  and the red lines correspond to $y = \tan{\beta}$,
  where $y$ is $\left. \left| \left( N_d \right)_{33} \right| \right/ m_b$ in (a)
  and $\left. \left| \left( N_u \right)_{33} \right| \right/ m_t$ in (b).}
\label{fig:Ntanb}
\end{figure}

The matrices $N_u$ and $N_d$ also exist in the usual flavour-preserving 2HDMs,
but there they are diagonal and proportional to the quark mass matrices.
In fact,
in the type-I 2HDM
\be
\label{u11}
N_u = - \frac{M_u}{\tan{\beta}}, \quad N_d = - \frac{M_d}{\tan{\beta}},
\ee
whereas in the type-II 2HDM
\be
\label{u22}
N_u = - \frac{M_u}{\tan{\beta}}, \quad N_d = M_d \tan{\beta}.
\ee
Now consider fig.~\ref{fig:Ntanb},
where we have plotted the values of both
$\left. \left| \left( N_d \right)_{33} \right| \right/ m_b$
and $\left. \left| \left( N_u \right)_{33} \right| \right/ m_t$
as functions of $\tan{\beta}$.
The green line shown in plot (a),
upon which many blue points are superimposed,
corresponds to $\left. \left| \left( N_d \right)_{33} \right| \right/ m_b
= 1 \left/ \tan{\beta} \right.$,
that one would obtain if the model behaved,
for the bottom quarks,
as a type-I 2HDM.
The red line would correspond to type-II behaviour,
\textit{viz.}\ $\left. \left| \left( N_d \right)_{33} \right| \right/ m_b
= \tan{\beta}$.
It appears that,
in this model,
most regions of parameter space yield either approximate type-I behaviour
or approximate type-II behaviour for bottom quarks.
Note that,
although the blue points appear superimposed on the green and red lines,
they are not {\em exactly}\/ on them---the type-I and type-II behaviours
displayed are {\em approximate}\/ and there are deviations from them,
which indeed can be large,
as we observe in particular for low values of $\tan{\beta}$.
In fig.~\ref{fig:Ntanb} (b) we observe the same behaviour
for top quarks---most points have either
$\left. \left| \left( N_u \right)_{33} \right| \right/ m_t \approx \tan{\beta}$
or $\left. \left| \left( N_u \right)_{33} \right| \right/ m_t \approx
\cot{\beta}$.

\begin{figure}
\begin{center}
\includegraphics[height=8cm,angle=0]{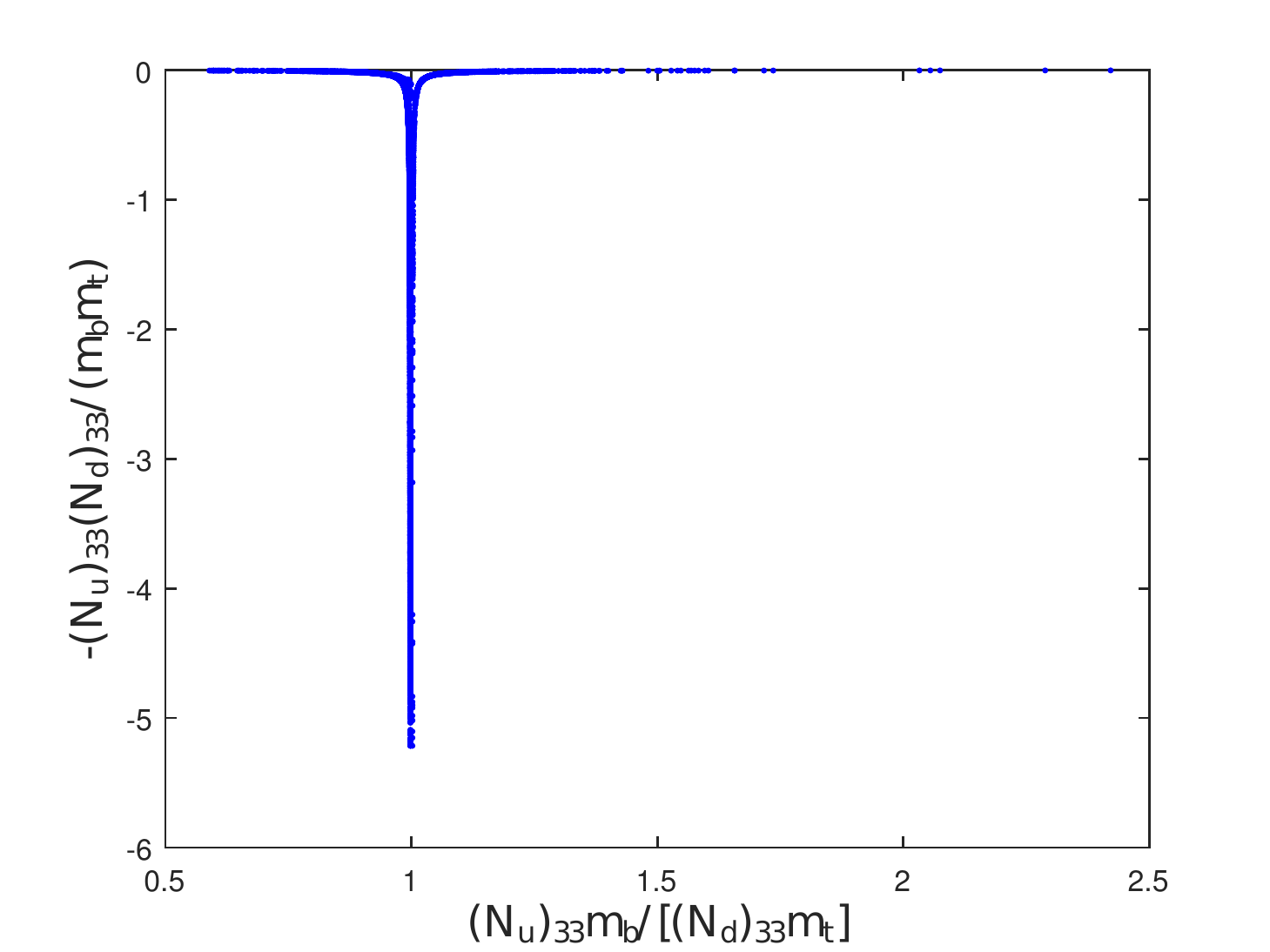}
\end{center}
\vspace{-5mm}
\caption{-$\left. \left( N_u \right)_{33} \left( N_d \right)_{33}
  \right/ \left( m_b m_t \right)$ \textit{versus}\/
  $\left. \left( N_u \right)_{33} m_b \right/
  \left[ \left( N_d \right)_{33} m_t \right]$.}
\label{figuraK}
\end{figure}

From equations~\eqref{u11} it follows that,
in the type~I 2HDM,
\be
\label{nunu1}
\frac{\left( N_u \right)_{33} m_b}{\left( N_d \right)_{33} m_t}
\ee
is equal to one,
whereas in the type~II 2HDM,
from equations~\eqref{u22},
\be
\label{nunu2}
- \frac{\left( N_u \right)_{33} \left( N_d \right)_{33}}{m_t m_b}
\ee
is equal to one.
In fig.~\ref{figuraK} we display the quantities~\eqref{nunu1} and~\eqref{nunu2}
plotted against each other.
That figure shows that,
for most points,
our model is more similar to the type-I 2HDM,
at least in what concerns giving mass to the third generation,
\textit{i.e.}\ the top-quark and bottom-quark masses originate mostly
in the Yukawa couplings to \emph{the same}\/ scalar doublet.
However,
there are also many allowed points for which
the quantity~\eqref{nunu1} is {\emph not}\/ unity;
for those points another regularity applies,
namely the quantity~\eqref{nunu2} is very close to zero.

Still,
one should remember that in our model there is flavour violation
in the Yukawa interactions,
and one obtains different results from those in
figs.~\ref{fig:Ntanb} and~\ref{figuraK}
for both the $\left( 1, 1 \right)$ and $\left( 2, 2 \right)$ entries
of both $N_d$ and $N_u$.
In fact,
the deviations from either type~I- or type~II-like behaviour
for the first and second generations are much more pronounced
than what one observes in fig.~\ref{fig:Ntanb}.
But,
the corresponding Yukawa couplings being much smaller,
that has much less importance for the Higgs-boson phenomenology
than the third-generation couplings that we have discussed in those figures.

\subsection{Properties of the extra scalars}

We now turn to the extra neutral scalars in the model,
$H$ and $A$.
The LHC Collaborations have been looking for neutral scalars
other than the 125\,GeV boson by investigating the production of $W^+W^-$,
$Z^0Z^0$,
and $\tau \bar \tau$,
among other channels.
The non-observation thus far of meaningful excesses in the cross sections,
relatively to their SM expectations,
imposes bounds on the masses and couplings of new particles.
In our model,
the imposition of the top- and meson-physics constraints
should force the off-diagonal entries of the matrices $N_d$ and $N_u$
to be smallish,
hence we expect that,
just as $h$,
the scalars $H$ and $A$ will have reduced flavour-changing interactions.
Therefore,
our model is expected to behave very much like the flavour-preserving 2HDMs
in what concerns the possibility of evading
the current experimental non-observation bounds for the extra scalars.
As we will now show,
there is a vast parameter space still allowed by the experimental constraints.

\begin{figure}
\begin{center}
\includegraphics[height=8cm,angle=0]{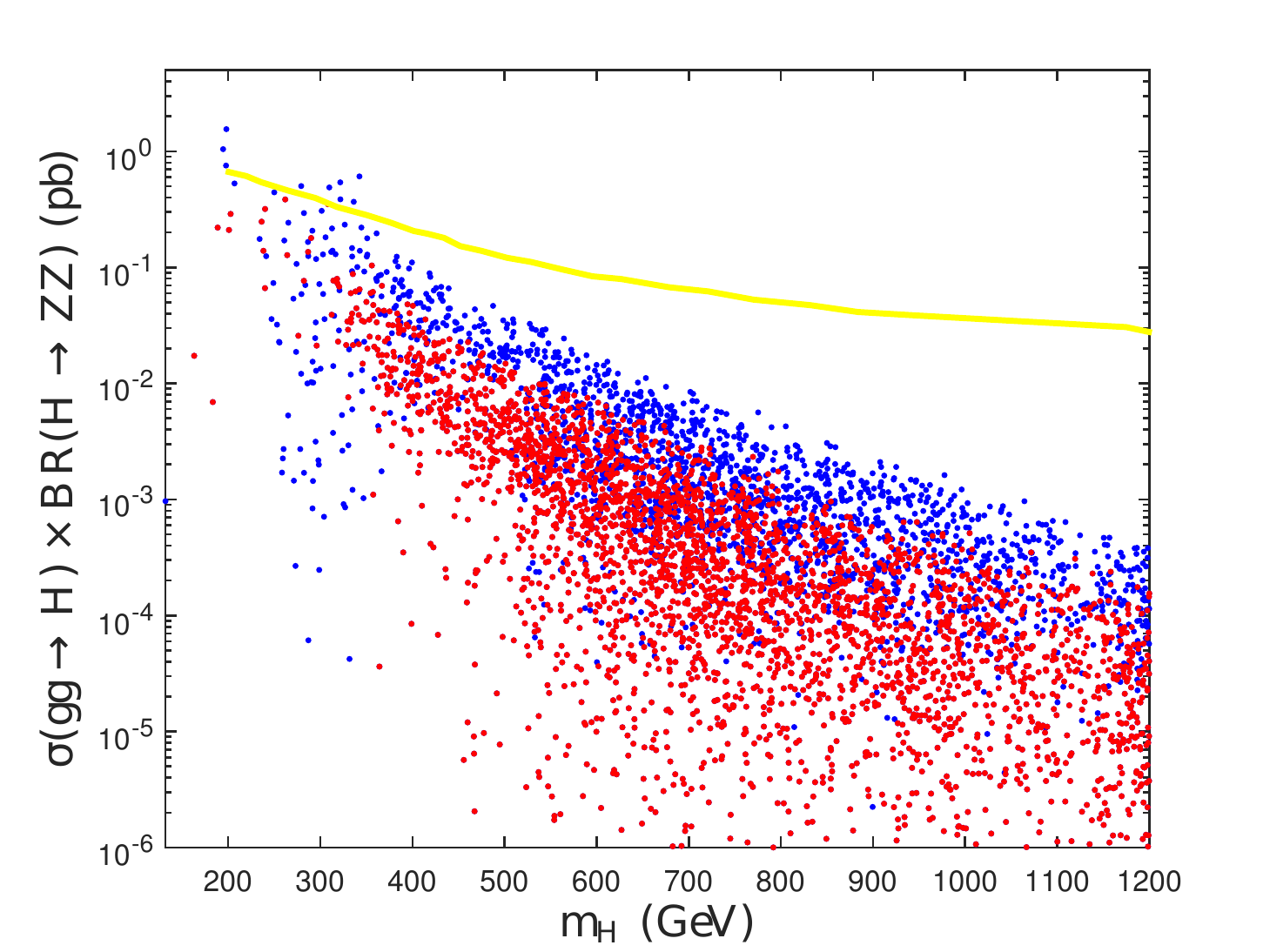}
\end{center}
  \vspace{-5mm}
\caption{Cross section of the process $pp \to gg \to H \to Z^0 Z^0$
  at 13\,TeV collision energy,
  as a function of the mass of the heavy $CP$-even scalar $H$.
  All the points displayed obey the constraints for our model
  described in subsection~\ref{sec:phen0}.
  For the blue points,
  the 125\,GeV scalar $h$ has all its production rates
  within 20\% of their SM-expected values;
  for the red points those production rates are all within 10\%
  of their SM values.
  The yellow line is the 2$\sigma$ upper bound
  given in ref.~\cite{ATLAS:2017spa}.}
\label{fig:HZZ}
\end{figure}

We firstly consider the limits coming from the search
for resonant $Z^0 Z^0$ pairs
by both the ATLAS~\cite{Aaboud:2017fgj,Aaboud:2017itg,
  Aaboud:2017gsl,ATLAS:2016oum,ATLAS:2017spa}
and CMS~\cite{CMS:2016noo,CMS:2016ilx,CMS:2017sbi} Collaborations.
This is a good channel to look for the heavy CP-even scalar $H$,
which may decay at tree level as $H \rightarrow Z^0 Z^0$.\footnote{Unlike
  the pseudoscalar $A$,
  which may decay to $Z^0 Z^0$ only through loops.}
In fig.~\ref{fig:HZZ} we show the points
which obey all the constraints described in section~\ref{sec:phen0};
the points in blue correspond to the requirement that
all the $\mu_X$ are within 20\% of 1,
and the points in red have all the $\mu_X$ less than 10\% away from 1.
The yellow line is the upper 2$\sigma$ bound
from the observed limit from ref.~\cite{ATLAS:2017spa}.
We observe that most of the allowed parameter space
yields a $pp \to gg \to H \to Z^0 Z^0$ cross section
below the experimental upper bound;
only a few low-$m_H$ points exceed the bound,
but even for those low values of $m_H$
there are plenty of points which are still allowed.
The tighter constraint of 10\% on the $h$ production rates
does not qualitatively change the picture.
There is a simple explanation for why low values
of the $pp \to gg \to H \to Z^0 Z^0$ event rate
should be obtained,
namely,
in any $CP$-conserving 2HDM
(or, indeed, multi-Higgs-doublet model)
there is the sum rule
\be
\label{vmlfdpd}
\left( g_{hZZ}^\mathrm{2HDM} \right)^2
+ \left( g_{HZZ}^\mathrm{2HDM} \right)^2
= \left( g_{hZZ}^\mathrm{SM} \right)^2
\ee
for the couplings of the $CP$-even neutral scalars to gauge-boson pairs.
Therefore,
if the coupling of $h$ to $Z^0$ (and $W^\pm$) pairs
is very close to its SM value,
then the coupling of $H$ to such pairs will be suppressed.
Equation~\eqref{vmlfdpd} is normally expressed through
$g_{hZZ}^\mathrm{2HDM} = s_{\beta - \alpha}\, g_{hZZ}^\mathrm{SM}$
and $g_{HZZ}^\mathrm{2HDM} = c_{\beta - \alpha}\, g_{hZZ}^\mathrm{SM}$;
SM-like behaviour of $h$ means $s_{\beta - \alpha} \simeq 1$,
which implies $c_{\beta - \alpha} \simeq 0$.

\begin{figure}[t]
\begin{tabular}{cc}
\includegraphics[height=6cm,angle=0]{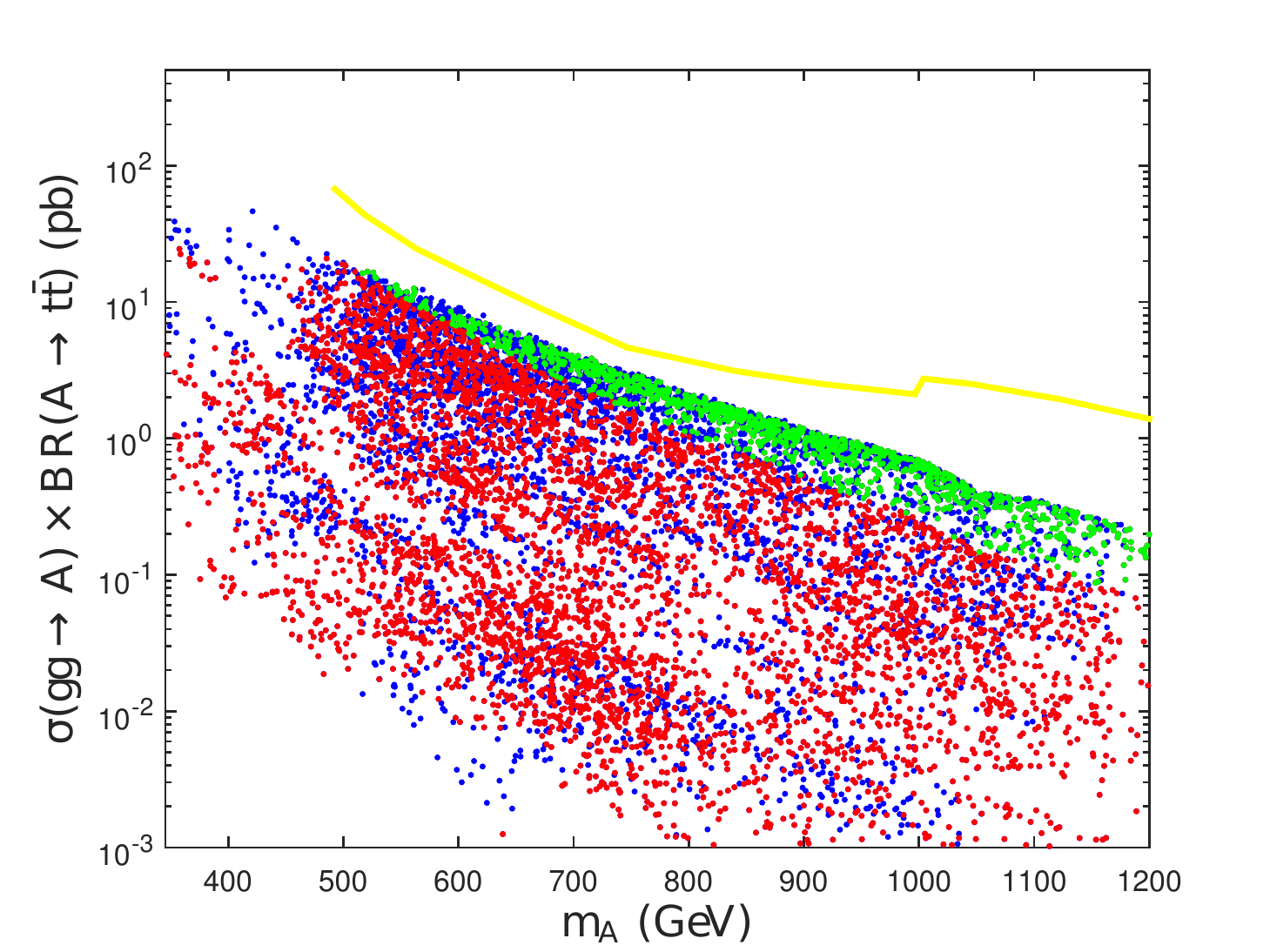}&
\includegraphics[height=6cm,angle=0]{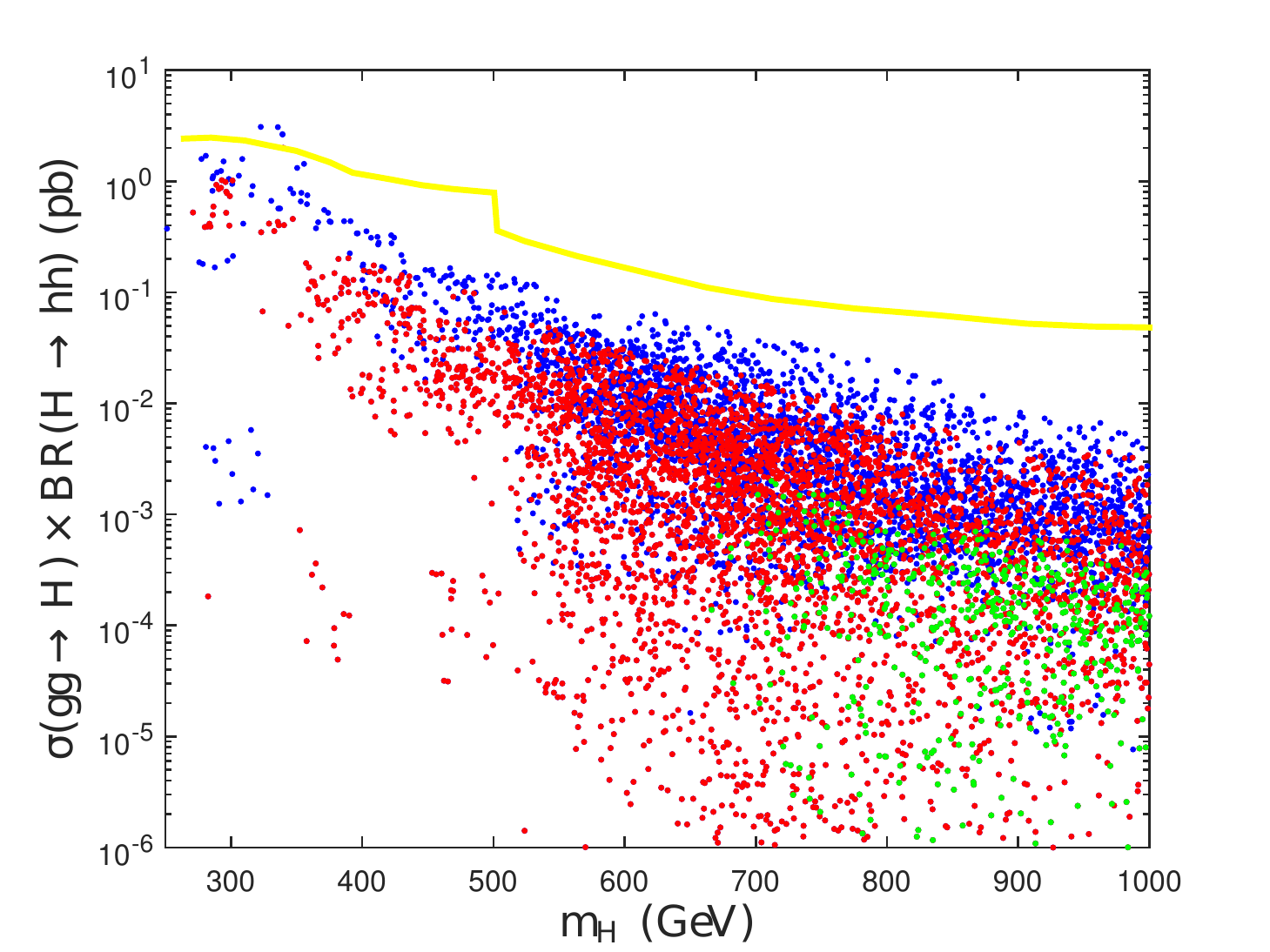}\\
 (a) & (b)
\end{tabular}
\caption{(a) The gluon--gluon production and decay to $t \bar t$
  of the pseudoscalar $A$ \textit{versus}\/ its mass.
  The yellow line is the upper 2$\sigma$ bound
  in figure 11 of ref.~\cite{Aaboud:2019roo}.
  (b) The gluon--gluon production and decay to $hh$
  of the scalar $H$ \textit{versus}\/ its mass.
  All the points displayed obey the constraints described
  in subsection~\ref{sec:phen}.
  For the blue points,
  $h$ has production rates within 20\% of its SM-expected values;
  for the red points those production rates are within 10\%
  of the SM value.
  The green points are a subset of the red ones,
  for which the width of the scalar in each plot is larger than 10\% its mass.
  The yellow line is the 2$\sigma$ upper bound
  in figure~6 of ref.~\cite{Aad:2015xja}.}
\label{fig:AH}
\end{figure}
In fig.~\ref{fig:AH} (a)
we show the gluon--gluon production cross section for a pseudoscalar $A$,
multiplied by its branching ratio to a $t \bar t$ pair,
at LHC.
(Similar results were obtained for $H$ instead of $A$,
but the obtained values of $\sigma \times \mathrm{BR}$
were about one order of magnitude lower than those of $A$.)
There are some LHC results for searches in the $t \bar t$
channel~\cite{Aaboud:2017hnm,Aaboud:2019roo};
we have used the results of ref.~\cite{Aaboud:2019roo},
although the analysis in that paper does not deal with the 2HDM.
In fig.~\ref{fig:AH} (a),
the yellow line is the upper 2$\sigma$ bound in figure 11
of ref.~\cite{Aaboud:2019roo}.\footnote{That figure concerns
  the possibility of a spin-2 Kaluza--Klein gravitation excitation,
  and it is the one for which the lowest values of
  $\sigma \times \mathrm{BR}$ are achieved,
  as well as the production channel where the initial state includes
  two gluons;
  we have chosen it as a conservative option.}
The published results only extend down to $m_A \simeq 500$\,GeV,
but it is clear that no exclusion will occur even for $A$ masses
lower than that.
As before,
the red (blue) points indicate a cut of 10\% (20\%)
on the $\mu_X$ ratios for the Higgs boson $h$,
meant to ensure its SM-like behaviour and compliance with the LHC results.
The green points in the same plot are the subset of the red ones
for which the width of $A$ is larger than 10\% of its mass:
$\left. \Gamma_A \right/ m_A > 0.1$.
We have thus far been assuming the validity of the narrow-width approximation
and neglecting eventual interferences between backgrounds and signal;
by marking these large-width points in green,
we want to draw attention to the only regions
where that approximation might fail.\footnote{Notice,
  though,
  that the width of $A$ is never larger than 29.2\%
  of its mass for the points obtained in our fit.}
The conclusion to draw from fig.~\ref{fig:AH} (a)
is that the current exclusion bounds from the $t \bar t$ resonance searches
are easily evaded by our model.

In fig.~\ref{fig:AH} (b)
we investigate the possibility of the heavy $CP$-even scalar $H$
being observed through its decay to two 125\,GeV scalars $h$.
This $hh$ channel is being thoroughly studied at the LHC,
considering several possible decay channels for both $h$
particles~\cite{Aad:2015xja,TheATLAScollaboration:2016ibb,
  CMS:2017ihs,Aaboud:2016xco,
  CMS:2017gxe,ATLAS:2016ixk,ATLAS:2016qmt,
  Sirunyan:2017guj,Sirunyan:2017djm};
the yellow line in fig.~\ref{fig:AH} (b) is the 2$\sigma$ upper bound
of figure~6 of ref.~\cite{Aad:2015xja}.
The blue and red points are the same as before;
the green points are the subset of the red ones
for which $\left. \Gamma_H \right/ m_H \geq 0.1$.
(Therefore,
the green points in fig.~\ref{fig:AH} (b)
do \emph{not}\/ coincide with the green points in fig.~\ref{fig:AH} (a).)
Unlike in fig.~\ref{fig:AH} (a),
the green points,
corresponding to scalars $H$ with a large width,\footnote{But,
  for all the points analyzed the width of $H$ was never larger than
  35\% of its mass.}
correspond to smaller values of $\sigma \times \mathrm{BR}$.
Just as in the previous figures,
we see that virtually all of our parameter space,
except a few low-mass points,
complies with the existing experimental bounds.

Since the model that we are studying differs from usual versions of the 2HDM
through the existence of FCNC,
we have considered the possibility of single-top decays
of the heavy (pseudo)scalars $H$ and $A$.
Indeed,
the non-diagonal Yukawa interactions
lead to the possibility of decays like $H \to t \bar u$ and $A \to c \bar t$,
which might be observed as top-quark + jet events at the LHC;
such events should be quite challenging to study in an hadronic machine
such as the LHC,
but the recent progress in charmed-jet identification algorithms
may be a significant contribution for a future analysis.\footnote{We thank
  Nikolaos Rompotis for this comment.}.
\begin{figure}[t]
\begin{tabular}{cc}
\includegraphics[height=6cm,angle=0]{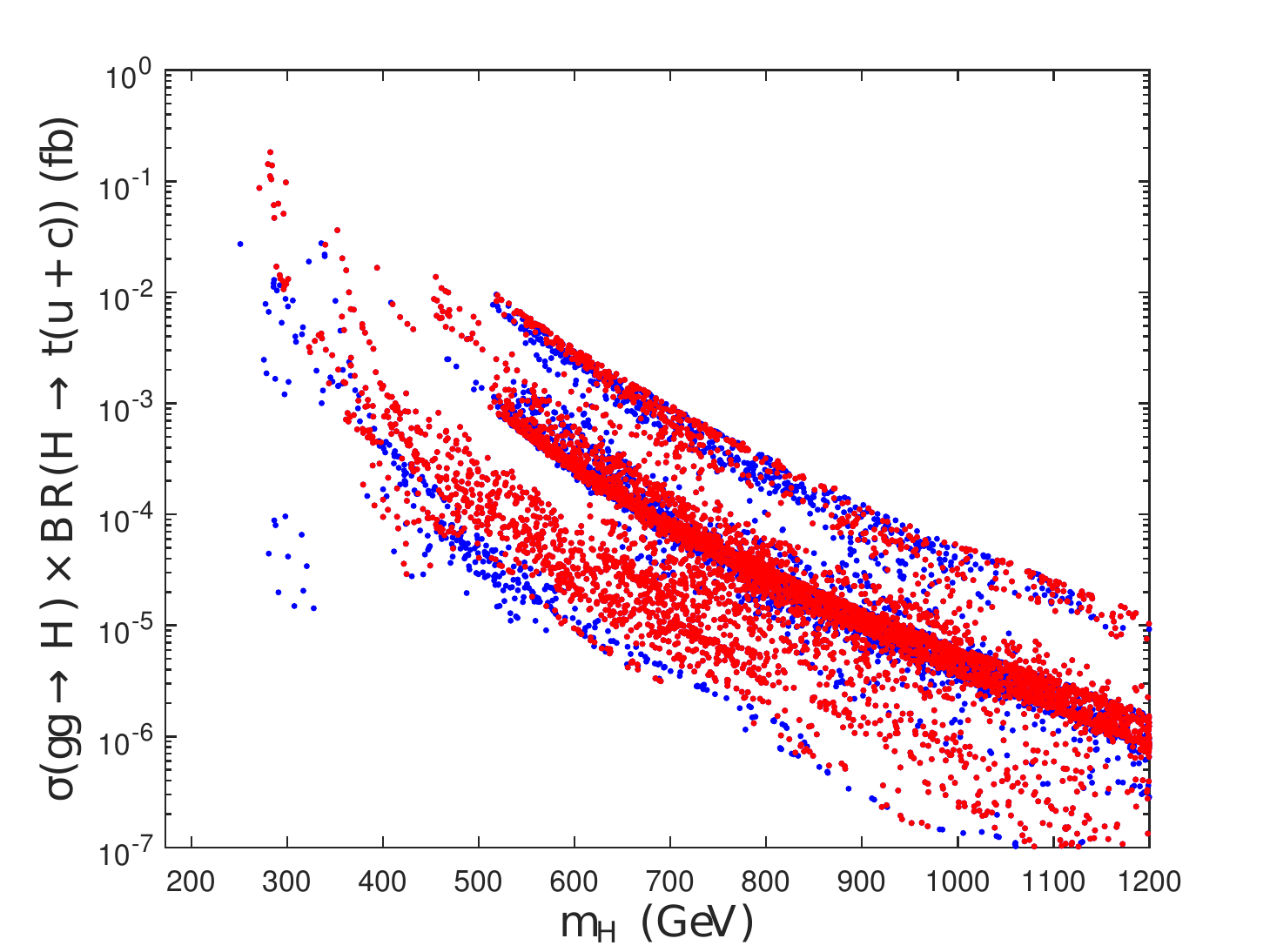}&
\includegraphics[height=6cm,angle=0]{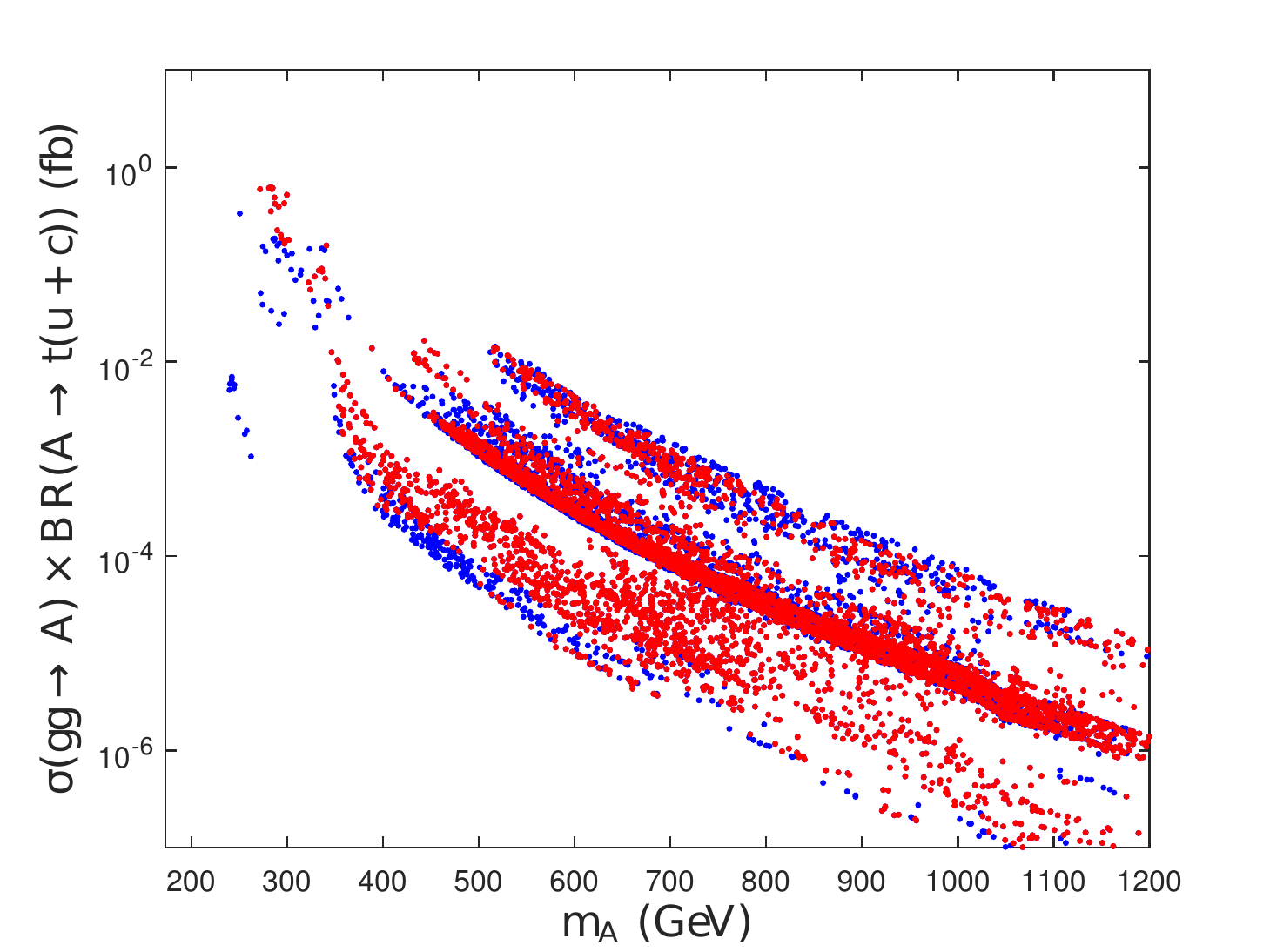}\\
 (a) & (b)
\end{tabular}
\caption{(a) The gluon--gluon production
  and the decay to $tq$ of (a) the pseudoscalar $A$
  and (b) the heavy $CP$-even scalar $H$
  \textit{versus}\/ their respective masses.
  The colour code is the same as in the previous figures.}
\label{fig:fcnc}
\end{figure}
In fig.~\ref{fig:fcnc} we present the expected cross section
times branching ratio for both (a) the $H$ and (b) the $A$.
In fig.~\ref{fig:fcnc} we have grouped together
all the FCNC decays of the scalars
with a single top in the final state,
\textit{viz.}\ the decays to $t \bar u$,
$\bar t u$,
$t \bar c$,
and $\bar t c$.
The top- and meson-physics constraints described in subsection~\ref{sec:phen0}
usually produce $N_u$ matrices with smallish off-diagonal elements,
and this yields very small branching ratios
for FCNC decays of both $H$ and $A$---the maximum values
that we have obtained were smaller than $5 \times 10^{-4}$,
but usual values were much smaller than that.
Therefore,
the model predicts values for $\sigma \times \mathrm{BR}$
usually orders of magnitude below the fentobarn.
It is difficult to find experimental bounds on such a search channel,
but the search for a $W^\prime$ decaying to a single top quark
plus a bottom quark~\cite{Aaboud:2018juj}
can at least give a rough idea of the current sensitivity of the LHC
for a top\,+\,jet resonance analysis.
Though the mass range is different
(the analysis of ref.~\cite{Aaboud:2018juj} starts at 1\,TeV),
the bounds shown in that paper
for the cross section times the branching ratio are of order 0.1\,pb,
and therefore much above the predicted $\sigma \times \mathrm{BR}$
shown for our model in fig.~\ref{fig:fcnc}.

\section{Conclusions}
\label{sec:conc}

In this paper we have presented a two-Higgs-doublet model that
attempts a partial solution of the strong $CP$ problem
by relegating a possible generation
of a nonzero $\theta$ to the two-loop level.
Our model achieves this by postulating a soft $CP$ violation
that transfers itself just to the CKM matrix,
with no $CP$ violation anywhere else in the model,
especially no $CP$ violation in scalar--pseudoscalar mixing.

We do not claim that our model achieves a full solution
of the strong $CP$ problem,
because the $\theta$ generated at two-loop level might still be too large.
However,
since in our model $CP$ violation exists only in the CKM matrix,
one may expect $\theta$ to be proportional to $J \sim 10^{-5}$,
the only $CP$-violating invariant quantity in that matrix.
Adding in a two-loop factor $\left( 16 \pi^2 \right)^{-2} \sim 10^{-4}$
and~\cite{ellis} probable suppression factors $m_q / m_W$,
where $m_q$ is a generic second-generation quark mass,
one might well reach a sufficiently small $\theta$.

Of course,
a 2HDM where $CP$ violation only occurs in the CKM matrix
eschews one of the motivations for multi-Higgs-doublet models,
namely,
obtaining extra sources of $CP$ violation
in order to reach a sufficiently large baryon number of the Universe.
We have nothing to say about this insufficiency.

We have investigated the compatibility of our model with the outstanding
experimental constraints,
in particular on
the flavour-changing neutral currents.
Our model can easily evade them,
at the price of cancelations
that might be qualified as fine-tuning.
We find that the new scalars in our model
may in some cases be little heavier than
the observed Higgs particle of mass 125\,GeV.
That will not necessarily make them easy to discover,
though, as we have seen in section~\ref{sec:phen}.

\paragraph{Acknowledgments:}
PF thanks Miguel Nebot for several enlightening discussions
concerning the fit to the meson sector, and Nuno Castro and Nikolaos Rompotis
for information concerning LHC experimental constraints.
PF is supported in part by the CERN fund grant CERN/FIS-PAR/0002/2017,
by the HARMONIA project under contract UMO-2015/18/M/ST2/00518
and by the CFTC-UL strategic project UID/FIS/00618/2019.
The work of LL is supported by the Portuguese
\textit{Funda\c c\~ao para a Ci\^encia e a Te\-cno\-lo\-gia}\/
through the projects PTDC/FIS-PAR/29436/2017,
CERN/FIS-PAR/0004/2017,
and UID/FIS/777/2013;
those projects are partly funded by POCTI (FEDER),
COMPETE,
QREN,
and the European Union.

\appendix

\section{The decay $\bar B \to X_s \gamma$}
\label{app:bsg}

The decays of a bottom-flavoured meson to a strange-flavoured meson
and a photon proceed via the quark transition $b \to s \gamma$.
Those decays constitute one of the most relevant constraints
on the parameter space of a multi-Higgs-doublet model,
because they receive important contributions from loops with charged scalars.
This is because the interactions of a charged scalar with down-type quarks
may be substantially enhanced by ratios of VEVs.
Consequently,
in a 2HDM the constraints from $b \to s \gamma$
typically eliminate substantial regions of the $m_{H^+}$--$\tan{\beta}$ plane.

In the model under discussion in this paper,
the occurrence of tree-level FCNC means that
the neutral scalars also contribute to $b \to s \gamma$,
unlike what happens in flavour-conserving 2HDMs.
We follow the general analysis of ref.~\cite{Lavoura:2003xp}
to take into account both the charged and the neutral scalars' contributions.
We write the Yukawa interactions of our model in the notation
\bs
\ba
\mathcal{L}_\mathrm{physical} &=& \cdots
+ \sum_{S = h, H, A} S \sum_{k, j = d, s, b} \bar k
\left( L^S_{kj} P_L + R^S_{kj} P_R \right) j
\\ & &
+ \left[ H^+ \sum_{\alpha = u, c, t} \bar \alpha \sum_{j = b, s, d}
\left( L^C_{\alpha j} P_L + R^C_{\alpha j} P_R \right) j
+ \mathrm{H.c.} \right],
\ea
\es
with coefficients defined as\footnote{In our model
  the matrices $N_u$ and $N_d$ are real,
  still we write the coefficients in the general form
  that follows from equation~\eqref{jvwo},
  \textit{viz.}\ allowing for complex $N_u$ and $N_d$.}
\bs
\ba
L^h_{kj} &=& \frac{s_{\beta - \alpha} m_k \delta_{kj}
  - c_{\beta - \alpha} \left( N_d^\ast \right)_{jk}}{v},
\\
R^h_{kj} &=& \frac{s_{\beta - \alpha} m_k \delta_{kj}
  - c_{\beta - \alpha} \left( N_d \right)_{kj}}{v},
\\
L^H_{kj} &=& \frac{c_{\beta - \alpha} m_k \delta_{kj}
  + s_{\beta - \alpha} \left( N_d^\ast \right)_{jk}}{v},
\\
R^H_{kj} &=& \frac{c_{\beta - \alpha} m_k \delta_{kj}
  + s_{\beta - \alpha} \left( N_d \right)_{kj}}{v},
\\
L^A_{kj} &=& \frac{i \left( N_d^\ast \right)_{jk}}{v},
\\
R^A_{kj} &=& \frac{- i \left( N_d \right)_{kj}}{v},
\\
L^C_{\alpha j} &=& \frac{\sqrt{2}}{v} \sum_{\beta = u, c, t}
\left( N_u^\ast \right)_{\beta \alpha} V_{\beta j},
\\
R^C_{\alpha j} &=& - \frac{\sqrt{2}}{v} \sum_{l = d, s, b}
V_{\alpha l} \left( N_d \right)_{l j}.
\ea
\es
The Wilson coefficients required for the computation of $b \to s \gamma$
are~\cite{Lavoura:2003xp}
\bs
\ba
C^\prime_7 \left( \mu \right) &=& g - \frac{e}{3}, \\
\Delta C_7 \left( \mu \right) &=& g^\prime - \frac{e^\prime}{3}, \\
C^\prime_8 \left( \mu \right) &=& f + e, \\
\Delta C_8 \left( \mu \right) &=& f^\prime + e^\prime,
\ea
\es
where
\bs
\ba
f &=& - \frac{v^2}{4 V_{ts}^\ast V_{tb} m_{H^+}^2} \sum_{\alpha = u, c, t}
{R^C_{\alpha s}}^\ast
\left[ R^C_{\alpha b}\, I_3 \left( \frac{m_\alpha^2}{m_{H^+}^2} \right)
  + L^C_{\alpha b}\, \frac{m_\alpha}{m_b}\,
  I_4 \left( \frac{m_\alpha^2}{m_{H^+}^2} \right) \right],
\label{j1} \\
g &=& - \frac{v^2}{4 V_{ts}^\ast V_{tb} m_{H^+}^2} \sum_{\alpha = u, c, t}
{R^C_{\alpha s}}^\ast
\left[ R^C_{\alpha b}\, I_5 \left( \frac{m_\alpha^2}{m_{H^+}^2} \right)
  + L^C_{\alpha b}\, \frac{m_\alpha}{m_b}\,
  I_6 \left( \frac{m_\alpha^2}{m_{H^+}^2} \right) \right],
\label{j2} \\
e &=& - \frac{v^2}{4 V_{ts}^\ast V_{tb}} \sum_{S = h, H, A}
\frac{1}{m_S^2} \sum_{k = d, s, b} {R^S_{ks}}^\ast
\left[ R^S_{kb}\, I_3 \left( \frac{m_k^2}{m_S^2} \right)
  + L^S_{kb}\, \frac{m_k}{m_b}\, I_4 \left( \frac{m_k^2}{m_S^2} \right) \right],
\ea
\es
and
\bs
\ba
f^\prime &=& f \left( R^C \leftrightarrow L^C \right),
\\*[2mm]
g^\prime &=& g \left( R^C \leftrightarrow L^C \right),
\\*[2mm]
e^\prime &=& e \left( R^S \leftrightarrow L^S \right)
\ \mbox{for} \ S = h, H, A.
\ea
\es
The functions $I_{3,4,5,6}$ are given in equations~(40)--(43)
of ref.~\cite{Lavoura:2003xp}.

To compute the overall branching ratio of $b \to s \gamma$
we follow refs.~\cite{Buras:1993xp,Misiak:2006ab}.
We use the effective operators described above,
defined as being at the Fermi scale $\mu = m_W$,
and include NLO QCD corrections by choosing $m_b$ as the renormalization scale.
Let $\eta =
\alpha_S \left( m_W \right) \left/ \alpha_S \left( m_b \right) \right.
= 0.5651$~\cite{Buras:1993xp} be the ratio of the running
strong coupling constant between scales $m_W$ and $m_b$.
We compute
\bs
\ba
\Delta C \left( m_b \right)
&=& \eta^{16/23}\, \Delta C_7 \left( m_W \right)
+ \frac{8}{3} \left( \eta^{14/23} - \eta^{16/23} \right)
\Delta C_8 \left( m_W \right),
\\
C^\prime \left( m_b \right) &=& \eta^{16/23}\, C^\prime_7 \left( m_W \right)
+ \frac{8}{3} \left( \eta^{14/23} - \eta^{16/23} \right)
C^\prime_8 \left( m_W \right),
\ea
\es
and then~\cite{Blanke:2012tv}
\bs
\ba
\mbox{BR} \left( b \to s \gamma \right) &=&
\mbox{BR} \left( b \to s \gamma\right)_\mathrm{SM}
\\ & & + \left( 2.47 \times 10^{-3} \right)
\left\{ \left| \Delta C \left( m_b \right) \right|^2
  + \left| C^\prime \left( m_b \right) \right|^2
  - 0.706\ \mbox{Re}{\left[ \Delta C \left( m_b \right) \right]} \right\},
  \hspace*{7mm}
\ea
\es
where $\mbox{BR} \left( b \to s \gamma \right)_\mathrm{SM} = 3.15 \times 10^{-4}$.
We consider our model to be in compliance with
the $b \to s \gamma$ data if it yields a branching ratio
within twice the experimental error bar,
\textit{viz.}\ we require $2.4406 \times 10^{-4} < \mbox{BR} \left( b \to s
\gamma \right)
< 3.8594 \times 10^{-4}$.

\section{The neutral meson--antimeson observables}
\label{app:mes}

The FCNC induced by the off-diagonal entries of $N_d$ and $N_u$
lead to tree-level contributions to flavour observables
such as $CP$ violation through the parameter $\epsilon_K$
and the mass differences in the $K^0$,
$B_d^0$,
$B_s^0$,
and $D^0$ meson--antimeson systems.
These are sensitive observables
and new-physics contributions to them may easily be overwhelming.
Thus,
we must make sure that the contributions to them
from the scalar sector of our model conform to the current data.
We use the numbers listed in ref.~\cite{Tanabashi:2018oca}.

\subsection{$K^0$--$\bar K^0$ observables}

Two $K^0$ meson observables are sensitive
to the tree-level FCNC contributions from the scalar sector:
the $CP$-violating parameter $\epsilon_K$ and
the mass difference between $K_S$ and $K_L$.
Both observables arise from the matrix element
effecting the trasition $\bar K^0 \to K^0$,
called $M_{21}$.
This receives contributions from the SM,
via box diagrams,
and from new physics (NP),
through FCNC in the scalar sector:
$M_{21} = M_{21}^\mathrm{SM} + M_{21}^\mathrm{NP}$.
We use the results presented in ref.~\cite{book}.
The SM contribution originates in a box diagram and is given by
\be
M_{21}^\mathrm{SM} = - \frac{G_F^2 m_W^2 f_K^2 m_K B_K}{12 \pi^2}
\left[ \eta_1\, \lambda_c^2\, S_0 \left( x_c \right)
  + \eta_2\, \lambda_t^2\, S_0 \left( x_t \right)
  + 2 \eta_3\, \lambda_c \lambda_t\, S_0 \left( x_c, x_t \right)
  \right],
\ee
where $G_F$ is the Fermi constant,
$m_W$ is the $W$-boson mass,
$f_K = 0.1555$\,GeV is the $K$-meson decay constant,
$m_K = 0.497611$\,GeV is the $K$-meson mass,
and $B_K = 0.723$ parameterizes the error in the vacuum-insertion approximation
for the relevant matrix element.
The $x_q = \left( m_q / m_W \right)^2$
and $\lambda_q = V_{qd}^\ast V_{qs}$ for $q = c, t$.
The functions $S_0$ are given in equations~(B.15) and~(B.16)
of ref.~\cite{book}.
Finally,
the parameters $\eta_1 = 1.38$,
$\eta_2 = 0.57$,
and $\eta_3 = 0.47$ account for QCD corrections.

The new-physics contribution originates in the tree-level exchange of $h$,
$H$,
and $A$.
One has,
by using the vacuum-insertion approximation
for the matrix elements of the operators,\footnote{See refs.~\cite{book}
  and~\cite{Ferreira:2011xc} for a detailed derivation of
  equation~\eqref{uhigfofp}.}
\bs
\label{uhigfofp}
\ba
M_{21}^\mathrm{NP}
&=& \frac{f_K^2 m_K}{96 v^2} \left\{
\left[ \left( N_d^\ast \right)_{ds}^2 + \left( N_d \right)_{sd}^2 \right]
\frac{10 m_K^2}{\left( m_s + m_d \right)^2}
\left( \frac{1}{m_A^2}
- \frac{c_{\beta-\alpha}^2}{m_h^2}
- \frac{s_{\beta-\alpha}^2}{m_H^2}
\right)
\right. \\ & & \left.
+ 4 \left( N_d^\ast \right)_{ds} \left( N_d \right)_{sd}
\left[ 1 + \frac{6 m_K^2}{\left( m_s + m_d \right)^2} \right]
\left( \frac{1}{m_A^2}
+ \frac{c_{\beta-\alpha}^2}{m_h^2}
+ \frac{s_{\beta-\alpha}^2}{m_H^2}
\right)
\right\}.
\label{eq:M12K}
\ea
\es
Notice that $M_{21}^\mathrm{NP}$ in our model is \emph{real}.

The $K_S$--$K_L$ mass difference is given by
$\Delta m_K = 2 \left| M_{21} \right|$.
Unfortunately,
the SM contribution to $\Delta m_K$ is affected by considerable uncertainties,
stemming from long-distance,
difficult to compute contributions to $M_{21}^\mathrm{SM}$,
and also from imprecisions in the value of $B_K$.
Therefore,
we just require that the new-physics term
does not give a contribution to $\Delta m_K$ larger than the experimental value,
\textit{i.e.}\ while fitting the parameters of the model we demand that
$2 \left| M_{21}^\mathrm{NP} \right| < 3.484 \times 10^{-15}$\,\mbox{GeV}.

It is expected that the uncertainties which trouble the calculation of
$\Delta m_K$ do not affect the computation of $\epsilon_K$,
given by
\be
\label{jfidsosp}
\epsilon_K = 2.228 \times 10^{-3} =
- \frac{\mbox{Im} \left( M_{21} {\lambda_u^\ast}^2 \right)}{\sqrt{2}\,
  \Delta m_K \left| \lambda_u \right|^2}.
\ee
In equation~\eqref{jfidsosp},
we use in the numerator $M_{21} = M_{21}^\mathrm{SM} + M_{21}^\mathrm{NP}$,
while for $\Delta m_K$ in the denominator we use the experimental value.
We accept fit results which give $\epsilon_K$ within a 10\% deviation
from the central value.

\subsection{$B_H$--$B_L$ mass differences}

The mass differences in the meson--antimeson $B_d^0$--$\bar B^0_d$
and $B_s^0$--$\bar B^0_s$ systems
are well measured and their theoretical calculation,
unlike that of $\Delta m_K$,
is reliable.
We use,
for the $\bar B^0_d \to B_d^0$ transition,
\be
\label{mfldp}
M_{21}^\mathrm{SM}\,=\,-\,\frac{G_F^2 m_W^2 f_{B_d}^2 m_{B_d} B_{B_d}}{12 \pi^2}\,
\eta_{B_d} \left( V_{tb} V^\ast_{td} \right)^2 S_0 \left( x_t \right),
\ee
with $f_{B_d} = 0.1902$\,GeV,
$m_{B_d} = 5.280$\,GeV,
$B_{B_d} = 1.219$,
and $\eta_{B_d} = 0.55$.
Note that in equation~\eqref{mfldp} one uses only
the box diagram with top-quark internal lines.
The NP contribution is given by an expression
analogous to equation~\eqref{uhigfofp},
with the obvious substitutions $f_K \rightarrow f_{B_d}$,
$m_K \rightarrow m_{B_d}$,
and $m_s \rightarrow m_b$.
We accept the result of the fit
if $2  \left| M_{21}^\mathrm{SM} + M_{21}^\mathrm{NP} \right|$
is within 10\% of the experimental value
$\Delta m_{B_d} = 3.333 \times 10^{-13}$\,GeV.

For the $\bar B_s^0 \to B^0_s$ transition
we have equation~\eqref{mfldp} with all indices $d \rightarrow s$
and $f_{B_s} = 0.228$\,GeV,
$m_{B_s} = 5.367$\,GeV,
$B_{B_s} = 1.28$,
and $\eta_{B_s} = 0.55$.
The NP contribution is given by an expression
analogous to eq.~\eqref{uhigfofp},
with the obvious substitutions $f_K \rightarrow f_{B_s}$,
$m_K \rightarrow m_{B_s}$,
and $m_d \rightarrow m_b$.
We accept the result of the fit
if $2 \left| M_{21}^\mathrm{SM} + M_{21}^\mathrm{NP} \right|$
is within 10\% of the experimental value
$\Delta m_{B_s} = 1.17 \times 10^{-11}$\,GeV.

\subsection{The mass difference in the $D^0$--$\bar D^0$ system}
\label{sec:tyufi}

There are also contributions
to the mass difference in the meson system $D^0$--$\bar D^0$.
As in the $K^0$--$\bar K^0$ system,
there are considerable uncertainties in the calculation
of $M_{21}^\mathrm{SM}$.
Therefore,
once again,
we resort to requiring only the New Physics contribution not to be too large.
We have
\bs
\ba
M_{21}^\mathrm{NP}
&=& \frac{f_D^2 m_D}{96 v^2} \left\{
\left[ \left( N_u^\ast \right)_{uc}^2 + \left( N_u \right)_{cu}^2 \right]
\frac{10 m_D^2}{\left( m_c + m_u \right)^2}
\left( \frac{1}{m_A^2}
- \frac{c_{\beta-\alpha}^2}{m_h^2}
- \frac{s_{\beta-\alpha}^2}{m_H^2}
\right)
\right. \\ & & \left.
+ 4 \left( N_u^\ast \right)_{cu} \left( N_u \right)_{uc}
\left[ 1 + \frac{6 m_D^2}{\left( m_c + m_u \right)^2} \right]
\left( \frac{1}{m_A^2}
+ \frac{c_{\beta-\alpha}^2}{m_h^2}
+ \frac{s_{\beta-\alpha}^2}{m_H^2}
\right) \right\},
\ea
\es
with $f_D = 0.212$\,GeV and $m_D = 1.865$\,GeV.
Conservatively,
we require $2 \left| M_{21}^\mathrm{NP} \right|$
to be smaller than the measured mass difference $6.253 \times 10^{-15}$\,GeV.

\section{$Z\rightarrow b\bar{b}$ constraints}
\label{app:Zbb}

A potentially very important constraint
to two-Higgs-doublet models (2HDMs)
stems from the measurement of the decay $Z \to b \bar b$.
We follow the treatment of that decay
in refs.~\cite{Haber:1999zh,Deschamps:2009rh,Degrassi:2010ne}.
The Lagrangian for the $Z b \bar b$ vertex is written as
\be
\mathcal{L}_{Z b \bar b} = - \frac{e Z_\mu}{s_W c_W}\, \bar b\, \gamma^\mu
\left( \bar{g}_b^L P_L + \bar{g}_b^R P_R \right) b,
\ee
where the coefficients $\bar{g}^{L,R}_b$ are,
at tree level in the SM,
$\bar{g}_b^L = - 1/2 + s^2_W/3$ and $\bar{g}_b^R = s^2_W/3$.
In both the SM and in extensions thereof,
these coefficients get one-loop contributions.
To wit,
in the 2HDM the contributions of loops with
charged scalars to $\bar{g}_b^L$ and $\bar{g}_b^R$
are given by
\bs
\ba
\delta\bar{g}_b^L &=& \frac{\sqrt{2} G_F}{16 \pi^2}
\left| \left( N_u^\dagger V \right)_{33} \right|^2
f_1 \left( \frac{m_t^2}{m^2_{H^+}} \right),
\\
\delta\bar{g}_b^R &=& -\frac{\sqrt{2} G_F}{16 \pi^2}
\left| \left( V N_d \right)_{33} \right|^2
f_1 \left( \frac{m_t^2}{m^2_{H^+}} \right),
\ea
\es
where
\be
f_1 \left( x \right) = \frac{x}{x - 1} \left( 1 - \frac{\ln{x}}{x - 1} \right).
\ee
The contributions of loops with neutral scalars are expected to be small,
both for 2HDMs with flavour
conservation~\cite{Haber:1999zh,Deschamps:2009rh,Degrassi:2010ne}
or without it~\cite{HernandezSanchez:2012eg};
we neglect them.
In order to take into account the current experimental results
on the observable quantities $R_b$ and $A_b$
(see refs.~\cite{Haber:1999zh,Tanabashi:2018oca}),
we have required that the charged-scalar contribution added to the SM one,
\textit{viz.}\ $\bar g^L_b = - 0.42112 + \delta \bar g^L_b$
and $\bar g^R_b = 0.07744 + \delta \bar g^R_b$,
does not deviate from the SM prediction by more than 2$\sigma$,
\textit{viz.}\ $2 \left( \bar g^L_b \right)^2
+ 2 \left( \bar g^R_b \right)^2
= 0.36782 \pm 0.00143$.

\section{A benchmark point}
\label{app:bench}

To illustrate the model,
we provide a specific point,
which is meant to serve only as an example.
The input in the scalar sector is
\bs
\label{jfodsddp}
\ba
& & v = \sqrt{v_1^2 + v_2^2} = 246\,\mathrm{GeV}, \quad
v_1 = 145.48\,\mathrm{GeV}, \quad
\beta - \alpha = 289.46^\circ,
\\
& & m_h = 125\,\mathrm{GeV}, \quad
m_H = 688.46\,\mathrm{GeV}, \quad
m_A = 364.01\,\mathrm{GeV}, \quad
m_{H^+} = 712.00\,\mathrm{GeV}. \hspace*{10mm}
\label{jvufidop}
\ea
\es
To this input correspond the following (approximate) values for the parameters
of the scalar potential:
\bs
\label{mbkflfp}
\ba
& & \mu_1 = 89682.55\,\mathrm{GeV}^2, \quad
\mu_2 = 31942.59\,\mathrm{GeV}^2, \quad
\mu_3 = 63188.85\,\mathrm{GeV}^2,
\\
& & \lambda_1 = 9.9708, \quad
\lambda_2 = 3.7121, \quad
\lambda_3 = 6.8336, \quad
\lambda_4 = -12.3750.
\ea
\es
It is easy to confirm that the values~\eqref{mbkflfp}
fulfil all the conditions~(\ref{eq:bfb}, \ref{condi2}, \ref{unitunit}).
Notice in equation~\eqref{jvufidop} that the masses of all four scalars
are neither too close nor too far away from each other.

The Yukawa-coupling matrices are as in equations~\eqref{cjviodp},
with
\bs
\ba
& &
b_1 = 1.0761 \times 10^{-3}, \quad
d_1 = 1.9555 \times 10^{-4}, \quad
f_1 = 5.1710 \times 10^{-5},
\\
&&
b_2 = -9.3709 \times 10^{-4}, \quad
d_2 = -3.0026 \times 10^{-2}, \quad
f_2 = 0,
\\
& &
p_1 = 6.9338 \times 10^{-2}, \quad
q_1 = -3.0282 \times 10^{-4}, \quad
r_1 = -1.3664 \times 10^{-2},
\\
& &
p_2 = -1.2295, \quad
q_2 = 0, \quad
r_2 = -9.2531 \times 10^{-3}.
\ea
\es
We also input $\aleph_1 = 0,\
\aleph_2 = 1.33\,\mathrm{rad}$.
One thus obtains quark masses and a CKM matrix in agreement
with equations~(\ref{quarkmasses}, \ref{ckmmatrix}).
The matrices that parameterize the FCNC are
\bs
\ba
N_d &=& \left( \begin{array}{ccccc}
  7.4 \times 10^{-3} & & 2.7 \times 10^{-2} & & 4.2 \times 10^{-2} \\
  -1.8 \times 10^{-5} & & 0.15 & & -5.0 \times 10^{-3} \\
  5.3 \times 10^{-5} & & -3.9 \times 10^{-7} & & -3.09
\end{array} \right),
\\*[1mm]
N_u &=& \left( \begin{array}{ccccc}
  1.2 \times 10^{-2} & & -0.29 & & -1.61 \\
  6.4 \times 10^{-2} & & -0.93 & & -14.9 \\
  -2.7 \times 10^{-3} & & 1.3 \times 10^{-4} & & -126
\end{array} \right).
\ea
\es
One sees that some off-diagonal matrix elements of $N_d$ are not very small,
and some off-diagonal matrix elements of $N_u$---which is almost
a triangular matrix---are pretty large.
There is a cancelation of about one part in 47
between the contributions to the neutral-$D$-meson mass difference
of the neutral scalars $h$ and $H$ on the one hand
and of the pseudoscalar $A$ on the other hand;
there are analogous,
yet milder,
cancelations among the contributions
to the other neutral-meson mass differences.
In general,
we have found that the $D$-meson mass difference constraint
requires quite strongly fine-tuned cancelations
when the neutral scalars have low masses,
while the constraints from all other neutral-meson systems
are much easier to satisfy and mostly require no fine-tuning.
Still,
notice that this benchmark point has one particle
(the pseudoscalar $A$)
with relatively low mass.

With this benchmark point,
the coupling modifiers defined
in equations~\eqref{eq:ktkb} and~\eqref{eq:ktau} are
\be
\kappa_V = 0.9993, \quad \kappa_t = 0.9721, \quad
\kappa_b = 0.9720, \quad \kappa_\tau = 1.0502,
\ee
and some of the phenomenological quantities computed in section~\ref{sec:phen}
are found to be:
\bs
\ba
\mathrm{BR} \left( h \to q\bar{q}^\prime \right) &=& 6.89\times 10^{-11}, \\
\sigma \left( gg \to h \right) &=& 38.98\,\mathrm{pb}, \\
\sigma \left( gg \to H \to Z^0Z^0 \right) &=& 0.75 \,\mathrm{fb}, \\
\sigma \left( gg \to A \to t \bar{q}_u \right) &=& 0.03 \,\mathrm{fb},
\ea
\es
where
(1) $q\bar{q}^\prime$ refers to a sum over all possible FCNC decays of $h$,
(2) the cross sections are for a LHC center-of-mass collision of 13\,TeV,
(3) the FCNC decays of $A$ involve $q_u = u$ and $q_u = c$,
and (4) we have grouped together all the FCNC decays
of the scalars with a single top in the final state.

\end{document}